\documentclass[12pt]{article}

\usepackage{epsfig}
\usepackage[latin1]{inputenc}
\usepackage{feynmf}

\newcommand{\beqn}{\begin{eqnarray}}
\newcommand{\eeqn}{\end{eqnarray}}

\def\lsim{\mathrel{\rlap{\lower4pt\hbox{\hskip1pt$\sim$}}
    \raise1pt\hbox{$<$}}}               
\def\gsim{\mathrel{\rlap{\lower4pt\hbox{\hskip1pt$\sim$}}
    \raise1pt\hbox{$>$}}}               

\newcommand{\<}{\langle}
\renewcommand{\>}{\rangle}
\newcommand{\GGt}{ {G \widetilde{G}} }
\newcommand{\D}{ \Delta }
\newcommand{\sla}{\hspace{-0.17cm}/}

\newcommand{\csw}{ c_{SW}}

\newcommand{\nn}{\nonumber}

\newcommand{\lm}[1]{\frac{\lambda^{#1}}{2}}

\newcommand{\ra}{\rightarrow}
\newcommand{\lra}{\longrightarrow}
\newcommand{\ms}{\overline{MS}}
\newcommand{\de}{\partial}

\newcommand{\mc}{\mathcal}
\newcommand{\tit}{\textit}
\newcommand{\scst}{\scriptstyle}

\newcommand{\NP}[4]{{#1}, Nucl. Phys. \textbf{#2}, {#3} ({#4}). }

\newcommand{\PL}[4]{{#1}, Phys. Lett. \textbf{#2}, {#3} ({#4}). }
\newcommand{\PR}[4]{{#1}, Phys. Rev. \textbf{#2}, {#3} ({#4}). }

\newcommand{\PRL}[4]{{#1}, Phys. Rev. Lett. \textbf{#2}, {#3} ({#4}). }

\begin{document}
\begin{fmffile}{diagsNPB03_WIs}

\rightline{\Large{Preprint RM3-TH/03-10}}

\vspace{1cm}

\begin{center}

\LARGE{Analysis of the Axial Anomaly on the Lattice with O(a)-improved Wilson Action}

\vspace{1cm}

\large{D. Guadagnoli$^1$ and S. Simula$^2$}

\vspace{0.5cm}

\normalsize{$^1$Dipartimento di Fisica, Universit\`a di Roma ``La Sapienza"\\
            Piazzale Aldo Moro 2, I-00185 Roma, Italy\\
            $^2$Istituto Nazionale di Fisica Nucleare, Sezione Roma III\\
            Via della Vasca Navale 84, I-00146 Roma, Italy}

\end{center}

\vspace{1cm}

\begin{abstract}

\noindent Flavor singlet and non-singlet axial Ward identities are investigated using the Wilson formulation of lattice $QCD$ with Clover $O(a)$-improvement, which breaks explicitly chiral symmetry. The matching at one-loop order of all the relevant renormalization constants with the continuum $\overline{MS}$ scheme is presented. Our calculations include: ~ 1) the contributions arising from the Clover term of the action; ~ 2) the complete mixing of the gluon operator $\GGt$ with the divergence of the singlet axial current; ~ 3) the use of both local and extended definitions of the fermionic bilinear operators. A definition of the gluon operator $\GGt$ on the lattice outside the chiral limit is proposed. Our definition takes into account the possible power-divergent mixing with the pseudoscalar density, generated by the breaking of chiral symmetry. A non-perturbative procedure for the evaluation of such mixing constant is worked out. Finally, the renormalization properties of the composite insertion of the topological charge operator $\int d^4x ~ \GGt(x)$ relevant for the lattice calculation of the neutron electric dipole moment, induced by the strong $CP$-violating term of the $QCD$ Lagrangian, are discussed.

\end{abstract}

\vspace{1cm}


\vspace{1cm}


\newpage

\pagestyle{plain}

\section{Introduction \label{section:introduction}}

\indent The investigation of the axial anomaly is a powerful tool to unravel the structure of the $QCD$ vacuum. Since the latter is highly non-trivial, the conservation of classical currents may be spoiled by quantum fluctuations of the vacuum. When this happens an anomaly is formed as in case of the flavor-singlet axial vector current. The axial anomaly can have deep consequences on various physical observables, like the large mass of the $\eta^{\prime}$ meson or the smallness of the flavor-singlet nucleon axial coupling constant, which is commonly known as the proton spin crisis.

\indent The connection between the topological structure of the $QCD$ vacuum and the axial anomaly has been elucidated by 't~Hooft \cite{topology}. A very interesting case where such an interrelation shows up, is provided by the neutron electric dipole moment ($EDM$) generated by the so-called $\theta$-term of the $QCD$ Lagrangian \cite{tHooft}, which in the Euclidean space is given by
 \beqn
    {\cal{L}}_{\theta} = i \theta {g^2 \over 32 \pi^2} \GGt = i\theta {g^2 
    \over 64 \pi^2} \varepsilon_{\alpha \beta \mu \nu} G_{\alpha \beta}^c
    G_{\mu \nu}^c ~ ,
    \label{eq:theta}
 \eeqn
where $G_{\mu \nu}^c$ is the gluon field strength, $c$ the color octet index ($c = 1, ..., 8$) and $\theta$ a dimensionless parameter. The $\theta$-term (\ref{eq:theta}) breaks both parity and time reversal symmetries and therefore it can generate a non-vanishing value of the neutron $EDM$\footnote{The present experimental upper limit on the neutron $EDM$ is $d_N \equiv |\vec{d}_N| < 6.3 \cdot 10^{-26} ~ (e \cdot cm)$ at $90 \%$ confidence level~\cite{ILL}, which corresponds to a severe bound on the magnitude of $\theta$. Indeed, using the available theoretical estimates from Refs.~\cite{Baluni,Crewther}, one has $d_N \approx 3 \cdot 10^{-16} ~ |\theta| ~ (e \cdot cm)$ leading to $|\theta| \lsim 2 \cdot 10^{-10}$. The smallness of the parameter $\theta$ is usually referred to as the strong $CP$ problem.}. Available estimates of the relevant matrix element however are based on phenomenological models, as the $MIT$ bag model of Ref.~\cite{Baluni} or as the effective $\pi N$ chiral Lagrangian of Ref.~\cite{Crewther}. Estimates relying on non-perturbative methods based on the fundamental theory, like lattice $QCD$, are still missing to date.

\indent In Ref.~\cite{EDM} a strategy for evaluating the neutron $EDM$ on the lattice induced by the strong $CP$ violating term (\ref{eq:theta}) was presented. Such a strategy is based on the standard definition of the neutron $EDM$, involving the insertion of the topological charge $(g^2 / 32 \pi^2) \int d^4x ~ \GGt(x)$ in the presence of the charge density operator $J_0$ (see Eq.~(4) of Ref.~\cite{EDM}). In case of three flavors with non-degenerate masses a complete diagrammatic analysis was performed \cite{EDM} showing how the axial anomaly governs the replacement of the topological charge operator with well-defined insertions of the flavor-singlet pseudoscalar density. The applicability of the method to the case of lattice formulations that break explicitly chiral symmetry, like the Wilson and Clover actions, was discussed in Ref.~\cite{EDM} using general arguments.

\indent The aim of this work is twofold: ~ i) to present a complete one-loop calculation of the renormalization constants appearing in both singlet and non-singlet axial Ward Identities ($WI$'s) using Wilson fermions with the Clover $O(a)$-improvement of the action; ~ ii) to provide a definition of the gluon operator $\GGt$ on the lattice outside the chiral limit, taking into account its possible power-divergent mixing with the pseudoscalar density. As for the one-loop matching, our calculations reproduce all the known results and add: ~ 1) the complete mixing of $\GGt$ with the divergence of the singlet axial current; and ~ 2) the use of both local and extended definitions of the fermionic bilinear operators. As for the calculation of the mixing between a lattice discretization of $\GGt$ and the pseudoscalar density in case of lattice formulations breaking chiral symmetry a non-perturbative procedure is presented. Finally, as a separate issue, the renormalization properties of the composite insertion of the topological charge operator $\int d^4x ~ \GGt(x)$ relevant for the lattice calculation of the neutron $EDM$ are discussed.

\indent The plan of the paper is as follows. In Section \ref{section:LWI} the structure of singlet and non-singlet axial $WI$'s using the Wilson and Clover lattice $QCD$ formulations is briefly recalled to fix notations and basic definitions. In Sections \ref{section:NSWIs} and \ref{section:SWIs} the non-singlet and singlet channels are considered, respectively. All the matching coefficients with the continuum $\overline{MS}$ scheme are explicitly calculated at one-loop order both with and without the $O(a)$-improvement of the action. The complete one-loop mixing of the gluon operator $\GGt$ with the divergence of the singlet axial current is evaluated. The issue of the possible power-divergent mixing of $\GGt$ with the pseudoscalar density is addressed and a non-perturbative procedure for evaluating the mixing constant is proposed. Moreover, in a separate subsection, the composite insertion of the topological charge operator $\int d^4x ~ \GGt(x)$ relevant for the lattice calculation of the neutron $EDM$ is considered and its renormalization properties are discussed. Section \ref{section:conclusions} is devoted to our conclusions. Finally, all the Feynman rules relevant for our calculations are collected in the Appendix.

\section{Axial Ward identities on the lattice \label{section:LWI}}

\indent In the subsequent discussion we will make use of the following definitions and notations. The $QCD$ action on the lattice is defined (in Euclidean space) as
 \beqn 
    S_{LQCD} = S_F + S_U
 \eeqn 
with $S_U$ being the pure Yang-Mills component \cite{Wilson} and $S_F$ the Wilson fermion action
 \beqn 
    S_F & = & a^4 \sum_{x} \Bigl\{ -\frac{1}{2a} 
    \sum_{\mu}[\bar{\psi}(x)(r-\gamma_\mu) U_\mu(x) \psi(x+\mu)
    \nn \\
    & + & \:\bar{\psi}(x+\mu)(r+\gamma_\mu)U_\mu^\dagger(x)\psi(x)] + 
    \bar{\psi}(x) \Bigl( m_0 + \frac{4r}{a}\Bigl)\psi(x) \Bigl\} ~,
    \label{S_F}
 \eeqn 
where color and flavor indices are omitted, $m_0$ is the (bare) mass matrix, diagonal in flavor, and the terms proportional to $r$ are necessary to avoid the fermion doubling. The improvement of the action is represented by the Clover term
 \beqn 
    S_C = - a^4 \sum_{x} \sum_{\mu, \nu} \:\csw \:\frac{i g_0 a r}{4} 
    \bar{\psi}(x)\sigma_{\mu \nu} P_{\mu \nu}(x) \psi(x)~,
 \eeqn 
with $P_{\mu \nu}$ being the usual lattice definition of the field-strength tensor $G_{\mu \nu}$ \cite{Gabrielli}
 \beqn 
    P_{\mu \nu}(x) = \frac{1}{4 a^2} \sum_{i=1}^4 \frac{1}{2 i g_0} (U_i - 
    U_i^\dagger)
    \label{Pmunu}
 \eeqn 
where the sum is over the four plaquettes in the $\mu$-$\nu$ plane stemming from $x$ and taken in the counterclockwise sense (see also Appendix \ref{feyrules}).

\indent To obtain axial $WI$'s on the lattice, one starts with the usual definition of the vacuum expectation value of an operator $\mc{O}(x_1, ..., x_n)$
 \beqn 
    \< \mc{O}(x_1, ..., x_n) \> = {1 \over Z_0} \int d[G] d[\psi]
    d[\bar{\psi}] ~ \mc{O}(x_1,...,x_n) ~ e^{-S}
    \label{vev}
 \eeqn 
where the fields are defined only on the nodes of the lattice, $Z_0$ is the partition function and $S = S_{LQCD} + S_C$. Performing local non-singlet axial rotations over the fermionic fields, namely
 \beqn 
   & & \psi(x) \rightarrow \left[ 1 + i \alpha^a(x) \lm{a} \gamma_5 \right]
   \psi(x) ~ , \nonumber \\[2mm]
   & & \bar{\psi}(x) \rightarrow \bar{\psi}(x) \left[ 1 + i \alpha^a(x)
   \lm{a} \gamma_5 \right] ~ ,
   \label{NS}
 \eeqn 
where $\lambda^a$ are the usual $SU(3)$ flavor matrices, and taking into account the invariance of the measure of integration in Eq.~(\ref{vev}), one gets
 \beqn 
    \< \mathcal{O} {\delta S \over \delta(i \alpha^a(x))} \> = \< {\delta
    \mathcal{O} \over \delta(i \alpha^a(x))} \>
    \label{NSWIa}
 \eeqn 
with
 \beqn 
    \< \mathcal{O} {\delta S \over \delta(i \alpha^a(x))} \> = - \D_\mu^x 
    \< \mc{O} A_{\mu}^a(x) \> + \< \mc{O} \bar{\psi}(x) \{\lm{a},m_0\} 
    \gamma_5 \psi(x) \> + \< \mc{O} \left[ X^a(x) + X_C^a(x) \right] \>~.
    \label{NSWIb}
 \eeqn 
In Eq.~(\ref{NSWIb}) we have indicated with $\D_\mu^x$ the backward derivative in the $\mu$-direction with respect to $x$. The non-singlet axial current $A_\mu^a$ is given by
 \beqn 
    A^{a}_\mu(x) = \frac{1}{2} \left[ \bar{\psi}(x) U_\mu(x) \gamma_\mu 
    \lm{a} \psi(x+\mu) + h.c. \right]~,
    \label{ANS}
 \eeqn 
and the operators $X^a$ and $X_C^a$ are the chiral variations of the Wilson and Clover term in the action, respectively,
 \beqn 
    X^a(x) & = & - \frac{r}{2a} \sum_{\mu} \Bigl[ \bar{\psi}(x) \lm{a} 
    \gamma_5 U_\mu(x) \psi(x+\mu) + \bar{\psi}(x+\mu) \lm{a} \gamma_5 
    U_\mu^\dagger(x) \psi(x) \nn \\
    & + & (x \rightarrow x-\mu) - 4 \bar{\psi}(x) \lm{a} \gamma_5 \psi(x) 
    \Bigl] ~,
    \label{Xa} \\[2mm]
    X_C^a(x) & = & - \frac{i g_0 a r}{2} \,\csw \sum_{\mu \nu} \bar{\psi}(x) 
    \lm{a} \gamma_5 \sigma_{\mu \nu} P_{\mu \nu}(x) \psi(x) ~.
    \label{XCa}
 \eeqn 

\indent Similarly, performing flavor-singlet rotations
 \beqn 
    \psi(x) & \rightarrow & \left[ 1 + i \alpha^0(x) \gamma_5 \right]
    \psi(x) ~ , \nonumber \\[2mm]
    \bar{\psi}(x) & \rightarrow & \bar{\psi}(x) \left[ 1 + i \alpha^0(x)
    \gamma_5 \right] ~ ,
   \label{S}
 \eeqn 
one obtains
 \beqn 
    \< \mathcal{O} {\delta S \over \delta(i \alpha^0(x))} \> = \< {\delta 
    \mathcal{O} \over \delta(i \alpha^0(x))} \>
    \label{SWIa}
 \eeqn 
where
 \beqn 
    \< \mathcal{O} {\delta S \over \delta(i \alpha^0(x))} \> = - \D^x_\mu 
    \<\mc{O} A_{\mu}(x) \> + 2 \< \mc{O} \bar{\psi}(x) ~ m_0 ~ \gamma_5 
    \psi(x) \> + \< \mc{O} \left[ X^0(x) + X^0_C(x)\right] \> ~.
    \label{SWIb}
 \eeqn 
The singlet axial current $A_\mu$ and the operators $X^0$ and $X^0_C$ are obtained from the corresponding octet operators (\ref{ANS})-(\ref{XCa}), respectively, with the na\"ive substitution $\lambda^a / 2 \ra 1$.

\indent Equations (\ref{NSWIa})-(\ref{NSWIb}) and (\ref{SWIa})-(\ref{SWIb}) are expressed in terms of unrenormalized quantities. The operators $X^a$, $X_C^a$, $X^0$ and $X_C^0$ are dimension-5 operators multiplied by one power of the lattice spacing; therefore, in any tree-level calculation they vanish identically in the continuum limit.

\indent With the inclusion of quantum corrections the situation changes drastically. In the non-singlet channel, the operators $X^a$ and $X_C^a$ mix with the axial current and with the pseudoscalar density: these mixings result in a finite (multiplicative) renormalization constant for the current $A_\mu^a$, and in an additive renormalization constant for the bare mass $m_0$, which multiplies the pseudoscalar density. The knowledge of such constants allows to identify the correct renormalized mass and axial current on the lattice, recalling that  Eqs.~(\ref{NSWIa})-(\ref{NSWIb}) should reproduce the corresponding continuum $WI$ in the limit $a \ra 0$.

\indent In the singlet channel, the mixings of the operator $X^0$ with the divergence of the (singlet) axial current and with the (singlet) pseudoscalar density are in general different from those corresponding to $X^a$, due to the presence of diagrams involving closed fermion loops (besides the ones present also in the non-singlet channel). This could cause in principle the singlet renormalized mass to be different from the non-singlet one: if it were so, we would be in trouble to identify the continuum limit of the lattice singlet $WI$, since in the continuum the renormalized quark masses in both the octet and the singlet channel are the same. However, using general non-perturbative arguments, it has been shown in Ref.~\cite{Testa} that renormalized masses are the same also on the lattice, so that, in this respect, a simple correspondence between the lattice $WI$'s and the continuum ones can be established.

\indent The singlet $WI$ on the lattice must reproduce the anomaly, represented by the term $2 N_F g_0^2/32 \pi^2 \GGt$, where $\GGt = 1/2 \,\varepsilon_{\mu \nu \rho \sigma} G^a_{\mu \nu} G^a_{\rho \sigma}$ and $G^a_{\mu \nu}$ is the usual gluon-field strength tensor. In a continuum regularization the singlet anomaly is generated by the fact that the regulator introduced during the renormalization process \tit{is not} chiral invariant. Removing the regulator leaves a residual contribution in the action proportional to $\GGt$.

\indent On the lattice, which is a regulator, the action (\ref{S_F}) with $m_0 = r = 0$ is perfectly chiral invariant, but it reproduces no anomaly, since it describes 16 quark species, with opposite chiral charges. A possible way to eliminate the 15 spurious fermions in the continuum limit, is represented by the Wilson term, which is however a chiral breaking term. The Wilson term generates the anomaly. Indeed, in the $WI$ (\ref{SWIa})-(\ref{SWIb}), the chiral variation of the Wilson term, namely $X^0(x)$, mixes with $\GGt$ in such a way to correctly reproduce the anomalous term present in the continuum~\cite{KarstenSmit}. The mixing is independent of $r$, as long as $r \neq 0$.

\indent Finally, given the anomalous non-conservation of the singlet axial current, the latter suffers an infinite (logarithmically divergent) renormalization. Analogous divergences occur for the operator $\GGt$, so that one needs a mixing between the two operators to end up with finite quantities. Furthermore, on the lattice with broken chiral symmetry, one has to take into account that the discretization adopted for $\GGt$ may mix with the pseudoscalar density.

\indent Since the $WI$'s (\ref{NSWIa})-(\ref{NSWIb}) and (\ref{SWIa})-(\ref{SWIb}) behave differently under renormalization, they will be treated separately in the next two Sections.

\section{Non-singlet axial $WI$'s} \label{section:NSWIs}

\indent In this Section we analyze in detail the one-loop structure of the $WI$ (\ref{NSWIa})-(\ref{NSWIb}). For easy of presentation we will consider the case of degenerate bare masses $m_0$, which means $\{ \lambda^a /2 , m_0\} \ra 2 m_0 \lambda^a /2$. First we notice that $X^a$ and $X_C^a$, having the same dimension and the same quantum numbers, should have an analogous behavior in perturbation theory. An analysis identical to the one made in Refs.~\cite{Bochicchio,Testa} for $X^a$ leads to the identification
 \beqn 
    X^a(x) + X_C^a(x) = \bar{X}^a(x) - 2 \bar{M}^{(NS)} \bar{\psi}(x) \lm{a} 
    \gamma_5 \psi(x) + \left( 1 - Z_A^{(NS)} \right) \de_\mu A_\mu^a(x) ~,
    \label{Xamix}
 \eeqn 
where the operator $\bar{X}^a(x)$ collectively refers to dimension-5 operators that vanish in the continuum limit for all matrix elements involving elementary operator insertions. Note that the renormalization constant $Z_A^{(NS)}$ refers to the extended definition of the non-singlet axial current, given in Eq.~(\ref{ANS}).

\indent The mixing coefficients $\bar{M}^{(NS)}$ and $[1 - Z_A^{(NS)}]$ appearing in Eq.~(\ref{Xamix}) can be computed at the one-loop level simply by considering the correlator $\< \mc{O}(x) \, \psi(y) \bar{\psi}(z) \>$ with amputation on the external legs, and $\mc{O}$ given by $X^a$ or $X^a_C$. Its Fourier transform can be pictorially represented by the following diagrams (the corresponding Feynman rules are reported in Appendix \ref{feyrules})
 \beqn
    \label{diagsXa}
    FT \< X^a(x) \, \psi(y) \bar{\psi}(z)  \>^{(2)}_{amp} &=& \quad \:\:
%
\parbox{5mm}{
  \begin{fmfgraph*}(15,15)
    \fmfleft{i1,i2} \fmfright{o1,o2}
    \fmf{plain,tension=3}{i1,v0} \fmfv{decor.shape=cross,decor.angle=55}{v0}
    \fmf{plain,tension=3}{v0,v1} 
    \fmf{quark}{v1,V1}
    \fmf{phantom}{i2,V1} \fmf{phantom}{V1,o2}
    \fmf{quark}{V1,v2}
    \fmf{plain,tension=3}{v2,v3} \fmfv{decor.shape=cross,decor.angle=-55}{v3}
    \fmf{plain,tension=3}{v3,o1}
    \fmffreeze
    \fmf{photon,right=0.2}{v1,v2}
    \fmfv{decor.shape=square,decor.size=4.thin,decor.filled=full}{V1} 
    \fmflabel{$\mc{O}$}{V1}
  \end{fmfgraph*}} \qquad \qquad \qquad \quad + 
%
 \parbox{5mm}{
  \begin{fmfgraph*}(15,15)  
    \fmfleft{i1,i2} \fmfright{o1,o2}
    \fmf{plain,tension=3}{i1,v0} \fmfv{decor.shape=cross,decor.angle=55}{v0}
    \fmf{plain,tension=3}{v0,v1}
    \fmf{quark}{v1,V1}
    \fmf{phantom}{i2,V1} \fmf{phantom}{V1,o2}
    \fmf{quark}{V1,v2}
    \fmf{plain,tension=3}{v2,v3} \fmfv{decor.shape=cross,decor.angle=-55}{v3}
    \fmf{plain,tension=3}{v3,o1}
    \fmffreeze
    \fmf{photon,right=0.6}{v1,V1}
    \fmfv{decor.shape=square,decor.size=4.thin,decor.filled=full}{V1} 
    \fmflabel{$\mathcal{O}$}{V1}
  \end{fmfgraph*} } \qquad \qquad \qquad \qquad \nn \\ 
 & + & \quad \:\:
%
\parbox{5mm}{
 \begin{fmfgraph*}(15,15)
   \fmfleft{i1,i2} \fmfright{o1,o2}
   \fmf{plain,tension=3}{i1,v0} \fmfv{decor.shape=cross,decor.angle=55}{v0}
   \fmf{plain,tension=3}{v0,v1}
   \fmf{quark}{v1,V1}
   \fmf{phantom}{i2,V1} \fmf{phantom}{V1,o2}
   \fmf{quark}{V1,v2}
   \fmf{plain,tension=3}{v2,v3} \fmfv{decor.shape=cross,decor.angle=-55}{v3}
   \fmf{plain,tension=3}{v3,o1}
   \fmffreeze
   \fmf{photon,right=0.6}{V1,v2}
   \fmfv{decor.shape=square,decor.size=4.thin,decor.filled=full}{V1} 
   \fmflabel{$\mathcal{O}$}{V1}
 \end{fmfgraph*} } \qquad \qquad \qquad \quad +
%
 \parbox{5mm}{
  \begin{fmfgraph*}(15,15)
    \fmfleft{i1,i2,i3} \fmfright{o1,o2,o3}
    \fmf{quark,tension=2}{i1,v0} \fmfv{decor.shape=cross,decor.angle=40}{v0}
    \fmf{plain,tension=2}{v0,V1}
    \fmf{phantom}{i2,V1} \fmf{phantom}{V1,o2}
    \fmf{plain,tension=2}{V1,v2} \fmfv{decor.shape=cross,decor.angle=-40}{v2} 
    \fmf{quark,tension=2}{v2,o1} 
    \fmffreeze
    \fmf{phantom}{i3,V2} \fmf{phantom}{V2,o3}
    \fmf{photon,right=1}{V1,V2} \fmf{photon,right=1}{V2,V1} 
    \fmf{phantom,tension=0.5}{V2,V1}
    \fmfv{decor.shape=square,decor.size=4.thin,decor.filled=full}{V1} 
    \fmflabel{$\mathcal{O}$}{V1}
  \end{fmfgraph*} } \qquad \qquad \qquad.
 \eeqn
One gets
 \beqn 
    \bar{M}^{(NS)} =\frac{g_0^2}{16 \pi^2} C_F  ( \frac{\Sigma_0}{a} + 
    m_0\,C) ~,
    \label{Mbar}
 \eeqn 
where
 \beqn 
    C & = & 16 {\pi}^2 \int_{-\pi}^{\pi} \frac{d^4 q}{(2\pi)^4} \Bigl\{ 
    \frac{2 r^2}{\D_2^2}(\D_1 \D_5 - \D_3) + \csw \frac{4 r^2}{\D_2^2} 
    (- \D_3 \D_5 + \D_4) \nn \\
    & + & \csw^2 \frac{2 r^2 \D_3}{\D_1 \D_2^2}(\D_3 \D_5 - \D_4) \Bigl\} 
    \label{C}
 \eeqn 
and $\Sigma_0$ is the additive mass renormalization, which appear in the one-loop approximation of the quark propagator [see below Eq.~(\ref{S0})].

\indent The one-loop expression for the (finite) multiplicative renormalization of the axial current is
 \beqn 
    Z_A^{(NS)} - 1 & = & \frac{g_0^2}{16 \pi^2} C_F \Bigl\{ 16 {\pi}^2 
   \int_{-\pi}^{\pi} \frac{d^4 q}{(2\pi)^4} r^2 \Bigl[ \frac{1}{2 \D_2} + 
   \frac{1}{\D_2^2}(5 \D_1 \D_3 + 4 \D_1 - 5 \D_1^2 +  \D_1^3 -13 \D_3 
   \nn \\
   & + & 4 \D_4) + \frac{r^2 \D_1}{\D_2^2}(\D_3 - 6 \D_1 + \D_1^2) +
   \csw \Bigl( \frac{1}{2 \D_1 \D_2}(4 \D_3 - \D_4) \nn \\
   & + & \frac{1}{\D_2^2}(10 \D_1 \D_3 - 2 \D_1 \D_4 -2 \D_1^2 \D_3 - 8 
   \D_3 - 4 \D_3^2 + 6 \D_4 - 4 \D_7) - \frac{\D_3}{2 \D_2} \Bigl) \nn \\
   & + & \csw^2 \Bigl( \frac{1}{\D_1 \D_2^2}(9 \D_3 \D_4 - 4 \D_3^2 - 2 
   \D_4^2) + \frac{1}{2 \D_1 \D_2}(4 \D_3 + \D_3^2 - 6 \D_4 + 2 \D_7 ) 
   \nn \\
   & + & \frac{1}{4 \D_2}(\D_1 \D_3 - 6 \D_3 + 3 \D_4 ) + 
   \frac{\D_3}{\D_2^2}(\D_3 - 2 \D_4)  \Bigl) + \csw \frac{2 r^2 
   \D_1}{\D_2^2}(\D_4  - \D_3 \D_5) \nn \\
   & + & \csw^2 \frac{r^2}{\D_2^2}\Bigl( \D_1 (-8 \D_3 - 3 \D_3^2 + 12 
   \D_4 - 4 \D_7 + 6 \D_1 \D_3 - 3 \D_1 \D_4 - \D_1^2 \D_3) \nn \\
   & - & \D_3 \D_4 + 4 \D_3^2 \Bigl) \Bigl] \Bigl\} ~.
   \label{Z_A}
 \eeqn 
The symbols $\D_i$ stand for the following functions
 \beqn 
    & & \D_1 = \sum_{\mu} \sin^2 \Bigl( \frac{q_\mu}{2}\Bigl) ~,\quad \quad
    \D_2 = \sum_{\mu} \sin^2(q_\mu) + \Bigl( 2 r \sum_{\mu} \sin^2 \Bigl( 
    \frac{q_\mu}{2}\Bigl)\Bigl)^2 ~, \nn \\
    & & \D_3 = \frac{1}{4} \sum_{\mu} \sin^2(q_\mu) ~,\quad \:\: \D_4 = 
    \sum_{\mu} \sin^2 \Bigl( \frac{q_\mu}{2}\Bigl) \cos^4 \Bigl( 
    \frac{q_\mu}{2}\Bigl) ~,\nn \\
    & & \D_5 = \sum_{\mu} \cos^2 \Bigl( \frac{q_\mu}{2}\Bigl) ~, \qquad 
    \qquad \qquad \D_6 = \sum_{\mu} \cos^4 \Bigl( \frac{q_\mu}{2}\Bigl) ~, 
    \nn \\
    & & \D_7 = \sum_{\mu} \sin^2 \Bigl( \frac{q_\mu}{2}\Bigl) \cos^6 \Bigl( 
    \frac{q_\mu}{2}\Bigl) ~ , \quad \quad ~ \D_8 = \sum_{\mu} \sin^2 \Bigl( 
    \frac{q_\mu}{2}\Bigl) \cos^8 \Bigl( \frac{q_\mu}{2}\Bigl) ~,
 \eeqn 
which for $\D_1, ..., \D_7$ are identical to the corresponding functions used in Ref.~\cite{Gabrielli}, while $\D_8$ is included here for completeness but used only in Section \ref{section:SWIs}. Numerical results for the constants $C$ and $Z_A^{(NS)}$ at various values of the Wilson parameter $r$ both for $\csw = 0$ and for $\csw = 1$ are reported in Table \ref{tab:CZ_A}.

\begin{table}[htb]

\vspace{0.5cm}

\begin{center}

\begin{tabular}{|*{5}{p{1.92cm}}|}
\hline \hline
$r$ &  $C$   & $Z_A^{(NS)}-1 $ & $C$ & $Z_A^{(NS)}-1$   \\
    &  $\scst \csw=0$   & $\scst \csw=0$ & $\scst \csw=1$ & $\scst \csw=1 $   \\ \hline
$0.0$  &0       &0                &0                  &0       \\
$0.2$  &8.09    &-3.62            &5.29               &-2.67   \\
$0.4$  &11.41   &-5.82            &6.20               &-4.13   \\
$0.6$  &11.25   &-7.00            &5.11               &-4.87   \\
$0.8$  &10.43   &-7.89            &3.96               &-5.37   \\
$1.0$  &9.64    &-8.66            &3.07               &-5.75   \\
\hline \hline

\end{tabular}

\end{center}

\caption{Numerical results for $C$ and $[Z_A^{(NS)} - 1]$ for different values of $r$ and $\csw$. A factor $g_0^2 C_F / 16 \pi^2$ is understood for $[Z_A^{(NS)} - 1]$ [see Eq.~(\ref{Z_A})].}

\label{tab:CZ_A}

\vspace{0.5cm}

\end{table}

\indent Equations (\ref{C})-(\ref{Z_A}), evaluated at $\csw = 0$, coincide with the ones obtained in Ref.~\cite{Bochicchio}, where the Clover term was not included. An explicit check of the correctness of the Clover contributions to Eqs.~(\ref{C}) and (\ref{Z_A}) can be obtained from the requirement that the $WI$ (\ref{NSWIa})-(\ref{NSWIb}) with $\mc{O} = \psi(y) \bar{\psi}(z)$ should be satisfied. This condition reads as
 \beqn 
    & & \D_\mu^x \< Z_A^{(NS)} A_\mu^a(x) \, \psi(y) \bar{\psi}(z) \> = 
    2 (m_0 - \bar{M}^{(NS)}) \< \bar{\psi}(x) \lm{a} \gamma_5 \psi(x) \, 
    \psi(y) \bar{\psi}(z) \> \nn \\
    & - & \delta(x-y) \< \left( \lm{a} \gamma_5 \psi(y) \right) \bar{\psi}(z) 
    \> - \delta(x-z) \< \psi(y) \left( \bar{\psi}(z) \lm{a} \gamma_5 \right) 
    \> + \< \bar{X}^a(x) \, \psi(y) \bar{\psi}(z)\> ~, ~~~~
    \label{check}
 \eeqn 
where we have taken into account Eq.~(\ref{Xamix}). In momentum space, denoting the amputated Green's functions with the insertion of the operators $\D_\mu A_\mu^a$ and $(\bar{\psi} \lambda^a \gamma_5 \psi / 2)$ by $\Lambda^a_{\D A_\mu}$ and $\Lambda^a_{\gamma_5}$, respectively, one gets
 \beqn 
    Z_A^{(NS)} \Lambda^a_{\D A^a}(p',p) = 2 \left(m_0 - \bar{M}^{(NS)} 
    \right) \Lambda^a_{\gamma_5}(p',p) - \gamma_5 \lm{a} S^{-1}(p) - 
    S^{-1}(p') \gamma_5 \lm{a} ~,
    \label{checkp}
 \eeqn 
where $p$ ($p'$) denotes momentum of the incoming (outcoming) quark, and $S(p)$ represents the quark propagator. The diagrams corresponding to $\Lambda^a_{\D A_\mu}$ are analogous to the ones considered for $X^a$ and $X^a_C$, reported in Eq.~(\ref{diagsXa}). For $\Lambda^a_{\gamma_5}$ one has to consider only the first diagram, being the pseudoscalar current a local operator. After calculating at one loop all the terms in Eq.~(\ref{checkp}), including the Clover contributions, one is able to check Eqs.~(\ref{Mbar})-(\ref{Z_A}). The explicit one-loop expression for the lattice inverse quark-propagator $S^{-1}(p)$ is
 \beqn 
    S^{-1}(p) = i p\sla + m_0 - \frac{{g_0}^2}{16 {\pi}^2} C_F \Bigl( 
    \frac{1}{a} \Sigma_0 \,+\, i  p\sla \Sigma_1(p) \,+\, m_0 \Sigma_2(p) 
    \Bigl)
 \eeqn 
with $\Sigma_{0,1,2}$ given by
 \beqn 
    \Sigma_0 & = &  16 {\pi}^2 \int_{-\pi}^{\pi} \frac{d^4 q}{(2\pi)^4} r 
    \Bigl \{ \frac{\D_3}{\D_1 \D_2} - \frac{1}{2 \D_1} -\frac{\D_5}{2 \D_2} 
    + r^2 \frac{\D_1}{2 \D_2} \nn \\
    & + & \csw \frac{1}{\D_1 \D_2}(\D_3 \D_5 - \D_4) + \csw^2 \frac{r^2}{2 
    \D_2}(\D_3 \D_5 - \D_4) \Bigl \} ~,
    \label{S0} \\[2mm]
    \Sigma_1(p) & = & \gamma_E - F_{0001} + 2 \int_0^1 dx \,x\ln \Bigl(a^2 
    m_0^2 (1-x) + a^2 p^2 x (1-x) \Bigl) \nn \\
    & + & 16 \pi^2 \int_{-\pi}^{\pi} \frac{d^4 q}{(2\pi)^4}  \Bigl\{ 
    \frac{\D_3}{16 \D_1^3} + \frac{1}{8 \D_1} - \frac{\D_3}{8 \D_1 \D_2} + 
    \frac{1}{8 \D_1^2 \D_2} (2 \D_4 - \D_3 \D_5) \nn \\
    & + & \frac{r^2}{4 \D_2}(2 - \D_1) + r^2 \csw \Bigl(\frac{1}{2 \D_1 
    \D_2}(\D_4 - 4 \D_3) + \frac{1}{8 \D_2}(- 9 \D_1 + 2 \D_1^2 \nn \\ 
    & + & 4 \D_3 +  4 \D_5 - 2 \D_6  + 4) \Bigl) + r^2 \csw^2 \Bigl( 
    \frac{1}{8 \D_1^2 \D_2}( - \D_3 \D_4 + 4 \D_3^2) \nn \\
    & + & \frac{1}{8 \D_1 \D_2} (-4 \D_3 \D_5 + 2 \D_3 \D_6  - 4 \D_3 + 4 
    \D_4 - \D_7) \nn \\ 
    & + & \frac{1}{8 \D_2}(-2 \D_1 \D_3 + 9 \D_3 - 2 \D_4) \Bigl)\Bigl\} ~,
    \label{S1} \\[2mm]
    \Sigma_2(p) & = & 4 \Bigl[\, \gamma_E - F_{0000} + \int_0^1 dx \, \ln 
    \Bigl( a^2 ~ m_0^2 (1-x) + a^2 ~ p^2 x (1-x) \Bigl) \Bigl] \nn \\
    & + & 16 \pi^2 \int_{-\pi}^{\pi} \frac{d^4 q}{(2\pi)^4} \Bigl\{ 
    - \frac{4 \D_1-\D_2}{4 \D_1^2 \D_2} + \frac{1}{4 \D_2} + 
    \frac{r^2}{\D_2^2}(2 \D_1 \D_5  + \frac{\D_2}{4} - 4 \D_3  ) \nn \\ 
    & + & r^4 \Bigl( -2 \frac{\D_1^2}{\D_2^2}\Bigl) + \csw \frac{4 
    r^2}{\D_2^2} (- \D_3 \D_5 + \D_4) +  \csw^2 \frac{r^2}{4 \D_1 
    \D_2}(\D_3 \D_5 - \D_4) \nn \\
    & + & \csw^2 \frac{2 r^4 \D_1}{\D_2^2}(-\D_3 \D_5 + \D_4) \Bigl\}
    \label{S2}
 \eeqn 
with $F_{0000} = 4.36898$, $F_{0001} = 1.31096$ and $\gamma_E = 0.577216$. Equations (\ref{S0}-\ref{S2}) coincide with the analytical results of Refs.~\cite{Bochicchio,Arroyo} (at $c_{SW} = 0$) and Ref.~\cite{Gabrielli} (at $c_{SW} = 1$).

\indent The check of Eqs.~(\ref{Mbar})-(\ref{Z_A}) via the $WI$ (\ref{checkp}) can proceed now through the projection of all the quantities onto the structures $(p \sla ' - p \sla) \gamma_5$ and $\gamma_5$ (orthogonal to each other) with the appropriate normalizations. In other words we calculate the quantities
 \beqn
    \Pi_{\Gamma}(p', p) = \frac{1}{N_\Gamma} Tr(\mc{T}(p',p) \Gamma) ~,
    \label{projected}
 \eeqn
where the trace is taken over the Dirac indices, $\mc{T}(p',p)$ stands for any of the terms appearing in Eq.~(\ref{checkp}), $\Gamma$ is equal to $(p \sla ' - p \sla) \gamma_5$ or $\gamma_5$ and $N_\Gamma$ is a normalization constant. We have calculated all the relevant projections and checked positively the correctness of Eqs.~(\ref{C})-(\ref{Z_A}).

\indent Note that the projection of the amputated diagram of the (extended) axial current over $(p \sla ' - p \sla) \gamma_5$ and of the pseudoscalar density over $\gamma_5$ can be compared with the corresponding quantities calculated in the continuum $\ms$-scheme in order to extract the matching constants between $\ms$ and the lattice  [see Subsect.~(\ref{ZmatchNS})], because the dependence on the external momenta is cancelled out in the difference between the two schemes. 

\indent Finally, from Eq.~(\ref{check}) it can be noticed that the explicit contribution of the operators $X^a$ and $X_C^a$ has been traded with a redefinition of the lattice axial current and of the bare mass $m_0$, i.e.
 \beqn 
    & & \D_\mu A^a_\mu \lra Z_A^{(NS)} \D_\mu A^a_\mu ~, \nn \\
    & & m_0 \lra m_0 - \bar{M}^{(NS)} \equiv m_L ~.
 \eeqn 

\subsection{Matching with the continuum in the non-singlet channel} \label{ZmatchNS}

\indent The Green's functions calculated with the (bare) fields and operators on the lattice can be matched to the corresponding ones renormalized in the $\ms$ scheme, by rescaling the fields and the operators on the lattice with appropriate constants. For instance the bare quark field on the lattice $\psi_L(x)$ can be rescaled through $\psi_L(x) = Z_\psi^{\frac{1}{2}} \psi_R(x)$, where the subscript $R$ indicates a renormalized ($\ms$) quantity in the continuum. The constant $Z_\psi$ can be evaluated via the relation 
 \beqn
    (S^{-1}(p))^R = Z_\psi (S^{-1}(p))^L
 \eeqn
which simply follows from the definition of the quark propagator. At one-loop order the relations
 \beqn
    & & (S^{-1}(p))^L = i p \sla \Bigl[ 1 - \frac{g_0^2}{16\pi^2} C_F 
    \Sigma_1^L(p) \Bigl] + m_L \Bigl[ 1 - \frac{g_0^2}{16\pi^2} C_F 
    \Sigma_2^L(p) \Bigl] ~, \nn \\
    & & (S^{-1}(p))^R = i p \sla \Bigl[ 1 - \frac{g_0^2}{16\pi^2} C_F 
    \Sigma_1^R(p) \Bigl] + m_R \Bigl[ 1 - \frac{g_0^2}{16\pi^2} C_F 
    \Sigma_2^R(p) \Bigl] ~,
    \label{propagators}
 \eeqn
imply 
 \beqn
    \label{DZpsi}
    Z_\psi = 1 - \frac{g_0^2}{16\pi^2}C_F \D_{\Sigma_1}, \qquad 
    \D_{\Sigma_1}= \Sigma_1^R(p) - \Sigma_1^L(p) ~.
 \eeqn
Analogously, one can define a matching for the mass on the lattice from the relation
 \beqn
    \label{defDZm}
    m_R = Z_m m_L,
 \eeqn
obtaining, thanks to the use of Eq.~(\ref{propagators}) and of the definition (\ref{DZpsi}),
 \beqn
    Z_m = Z_\psi (1 - \frac{g_0^2}{16\pi^2}C_F \D_{\Sigma_2}) ~.
    \label{DZ_m}
 \eeqn

\indent A similar procedure leads to the definition of the $Z$'s for the matching of composite operators. In the non-singlet channel, we are interested in Green's functions involving fermion bilinear operators $\mc{O}$ contracted with two external quark fields, with amputation on the external legs. The corresponding functions in momentum space will be indicated by $\Lambda_\mc{O}(p', p)$. One gets the relation
 \beqn
    \Lambda_{\mathcal{O}}^R(p', p) = Z_\psi Z_{\mc{O}} 
    \Lambda_{\mathcal{O}}^L(p', p)
    \label{defZmatchO}
 \eeqn
which defines $Z_{\mc{O}}$ as the constant of matching for the operator $\mc{O}$. In perturbation theory one can express the constants in Eq.~(\ref{defZmatchO}) as
 \beqn
    \label{ZmatchO2}
    Z_\psi Z_{\mc{O}} = 1 + \frac{g_0^2}{16\pi^2}C_F \D_{\mc{O}} ~,
    \qquad \D_{\mc{O}} = \hat{\Lambda}_{\mc{O}}^R - \hat{\Lambda}_{\mc{O}}^L ~,
 \eeqn
where the expressions $\hat{\Lambda}$ collect all the one-loop diagrams. The relation (\ref{ZmatchO2}) implies
 \beqn 
    \label{DZO=}
    Z_{\mc{O}} = 1 + \frac{g_0^2}{16\pi^2}C_F \Bigl( \D_{\mc{O}} + 
    \D_{\Sigma_1} \Bigl)~.
 \eeqn

\indent Denoting the subtraction scale in the continuum by the symbol $\mu$, our results for ${\D}_{\Sigma_1}$ and ${\D}_{\Sigma_2}$ are
 \beqn 
    \label{DZS1}
    {\D}_{\Sigma_1} & = & 1 - \ln a^2 \mu^2 -\gamma_E + F_{0001} - 16 \pi^2 
    \int_{-\pi}^{\pi} \frac{d^4 q}{(2\pi)^4}  \Bigl\{ \frac{\D_3}{16 \D_1^3} 
    + \frac{1}{8 \D_1} - \frac{\D_3}{8 \D_1 \D_2} \nn \\
    & + & \frac{1}{8 \D_1^2 \D_2} (2 \D_4 - \D_3 \D_5) + \frac{r^2}{4 
    \D_2}(2 - \D_1) + r^2 \csw \Bigl(\frac{1}{2 \D_1 \D_2}(\D_4 - 4 \D_3)
    \nn \\
    & + & \frac{1}{8 \D_2}(- 9 \D_1 + 2 \D_1^2  + 4 \D_3 +  4 \D_5 - 2 \D_6  
    + 4) \Bigl) \nn \\
    & + & r^2 \csw^2 \Bigl( \frac{1}{8 \D_1^2 \D_2}( - \D_3 \D_4 + 4 \D_3^2)
    + \frac{1}{8 \D_1 \D_2} (-4 \D_3 \D_5 + 2 \D_3 \D_6  \nn \\
    & - & 4 \D_3 + 4 \D_4 - \D_7) + \frac{1}{8 \D_2}(-2 \D_1 \D_3 + 9 \D_3 - 
    2 \D_4) \Bigl) \Bigl\}, \quad \:
 \eeqn
\medskip
 \beqn
    \label{DZS2}
    {\D}_{\Sigma_2} & = & 4 \Bigl(\, \frac{1}{2} - \ln a^2 \mu^2 -\gamma_E 
    + F_{0000} \Bigl) - 16 \pi^2 \int_{-\pi}^{\pi} \frac{d^4 q}{(2\pi)^4} 
    \Bigl\{ -\frac{4 \D_1-\D_2}{4 \D_1^2 \D_2} + \frac{1}{4 \D_2} 
    \nn \\
    & + & \frac{r^2}{\D_2^2}(2 \D_1 \D_5  + \frac{\D_2}{4} - 4 \D_3) 
    + r^4 \Bigl( -2 \frac{\D_1^2}{\D_2^2}\Bigl) \nn \\ 
    & + & \csw\frac{4 r^2 }{\D_2^2} (- \D_3 \D_5 + \D_4) + \csw^2 
    \frac{r^2}{4 \D_1 \D_2}(\D_3 \D_5 - \D_4) + \nn \\ 
    & + & \csw^2 \frac{2 r^4 \D_1}{\D_2^2}(-\D_3 \D_5 + \D_4) \Bigl\} ~,
 \eeqn

\indent The numerical results obtained for Eqs.~(\ref{S0}), (\ref{DZS1}) and (\ref{DZS2}) at various values of $r$, are reported in Tables \ref{tab:S_csw=0} and \ref{tab:S_csw=1} for the two cases $\csw = 0$ and $\csw = 1$, respectively.

\begin{table}[htb]

\begin{center}

\begin{tabular}{|*{4}{p{2.4cm}}|}
\hline \hline
$r$ & $\Sigma_0$ &  $\D_{\Sigma_1}$   & $\D_{\Sigma_2}$  \\
\hline
$0.0$  &0                &-6.04            &33.17       \\
$0.2$  &-19.79           &-7.13            &18.63       \\
$0.4$  &-30.70           &-8.91            &8.35        \\
$0.6$  &-38.29           &-10.47           &3.86        \\
$0.8$  &-44.96           &-11.77           &1.55        \\
$1.0$  &-51.43           &-12.85           &0.10        \\
\hline \hline

\end{tabular}

\end{center}

\caption{Numerical results for $\Sigma_0$ and $\D_{\Sigma_{1,2}}$ [see Eqs.~(\ref{S0}), (\ref{DZS1}) and (\ref{DZS2})] for different values of $r$ and with $\csw=0$. The choice $\mu = 1/a$ is understood.}

\label{tab:S_csw=0}

\end{table}

\begin{table}[htb]

\vspace{0.5cm}

\begin{center}

\begin{tabular}{|*{4}{p{2.4cm}}|}
\hline \hline
$r$ & $\Sigma_0$ &  $\D_{\Sigma_1}$   & $\D_{\Sigma_2}$ \\
\hline
$0.0$  &0       &-6.04            &33.17              \\
$0.2$  &-12.04  &-6.52            &21.81              \\
$0.4$  &-18.31  &-7.40            &14.69              \\
$0.6$  &-22.98  &-8.16            &11.95              \\
$0.8$  &-27.44  &-8.75            &10.74              \\
$1.0$  &-31.99  &-9.21            &10.10              \\
\hline \hline

\end{tabular}

\end{center}

\caption{The same as in Table \ref{tab:S_csw=0} but with $\csw=1$.}

\label{tab:S_csw=1}

\vspace{0.5cm}

\end{table}

\indent As for the quantities $\D_I$, $\D_{\gamma_5}$ and $\D^{ext}_{\gamma_\mu \gamma_5}$ one obtains
 \beqn 
    \label{DZI}
    \D_I & = &  4 \Bigl\{ -1 + \ln a^2 \mu^2 - F_{0001} + \gamma_E - 16 
    \pi^2 \int_{-\pi}^\pi \frac{d^4q}{(2 \pi)^4} \Bigl[ - \frac{\D_3}{16 
    \D_1^3} + \frac{\D_3 \D_5}{4 \D_1 \D_2^2} \nn \\
    & + & \frac{r^2}{4 \D_2^2}(-\D_1 \D_5 + 3 \D_3) + r^4 \frac{\D_1^2}{4 
    \D_2^2} + \csw \frac{r^2}{\D_2^2} (\D_3 \D_5 - \D_4) \nn \\
    & + & \csw^2 \frac{r^2 \D_3}{4 \D_1 \D_2^2}(-\D_3 \D_5 + \D_4) + \csw^2 
    \frac{r^4 \D_1}{4 \D_2^2} (\D_3 \D_5 - \D_4) \Bigl] \Bigl\} ~, \:\:\:\:
 \eeqn
 \beqn
    \label{DZg5}
    \D_{\gamma_5} & = & 4 \Bigl\{ -1 + \ln a^2 \mu^2 - F_{0001} + \gamma_E
    - 16 \pi^2 \int_{-\pi}^\pi \frac{d^4q}{(2 \pi)^4} \Bigl[ - \frac{\D_3}{16 
    \D_1^3} + \frac{\D_3 \D_5}{4 \D_1 \D_2^2} \nn \\
    & + & \frac{r^2}{4 \D_2^2}(\D_1 \D_5 +  \D_3) + r^4 \frac{\D_1^2}{4 
    \D_2^2} + \csw^2 \frac{r^2 \D_3}{4 \D_1 \D_2^2}(\D_3 \D_5 - \D_4) \nn \\
    & + & \csw^2 \frac{r^4 \D_1}{4 \D_2^2} (\D_3 \D_5 - \D_4) \Bigl] \Bigl\} 
    ~, \:\:\:\:
 \eeqn
 \beqn
    \label{DZgmug5ext}
    \D^{ext}_{\gamma_\mu \gamma_5} & = & \Pi^{(2) R}_{J^5_\mu} - 
    \Pi^{(2)}_{J^{5,ext}_\mu} = \nn \\ 
    & = & - 2 + \ln a^2 \mu^2 - F_{0001} + \gamma_E - 16 \pi^2 \int_{-\pi}^\pi 
    \frac{d^4q}{(2 \pi)^4} \Bigl[ - \frac{\D_3}{16 \D_1^3} + \frac{\D_3}{\D_1 
    \D_2^2} - \frac{1}{8 \D_1} \nn \\
    & + & \frac{\D_3}{4 \D_1 \D_2} + \frac{1}{\D_1 \D_2^2}(3 \D_3 + \D_3^2 - 
    5 \D_4 + 2 \D_7) + \frac{1}{2 \D_2^2}(\D_1 \D_3 - 6 \D_3 \nn \\
    & + & 2 \D_4) + r^2\frac{\D_1}{4 \D_2} + \frac{r^2}{2 \D_2^2}( -3 \D_1 
    \D_3 - 4 \D_1 \D_5 + 2 \D_1 \D_6 + 4 \D_1 + 3 \D_1^2 \nn \\
    & - & \D_1^3 + 7 \D_3 - 2 \D_4) + \frac{r^4 \D_1^2}{2 \D_2^2}(2 - \D_1) 
    + \csw \frac{r^2}{\D_2^2}(-7 \D_1 \D_3 + 4 \D_1 \D_4 \nn \\
    & + & 2 \D_1^2 \D_3 + 8 \D_3 \D_5 - 4 \D_3 \D_6 - 4 \D_3 - 15 \D_4 + 6 
    \D_7) \nn \\ 
    & + & \csw^2 \frac{r^2}{2 \D_1 \D_2^2} (3 \D_3 \D_4 - 4 \D_3 \D_7 - 4 
    \D_3^2 \D_5 + 2 \D_3^2 \D_6 + 4 \D_3^2 + 2 \D_4^2 \nn \\ 
    & + & \D_1 (3 \D_3^2 - \D_3 \D_4 - \D_1 \D_3^2)) + \csw^2 \frac{r^4 
    \D_1}{2 \D_2^2}(4 \D_3 \D_5 - 2 \D_3 \D_6 \nn \\ 
    & - & 12 \D_4 + 4 \D_7 - 4 \D_1 \D_3 + 3 \D_1 \D_4 + \D_1^2 \D_3 ) \Bigl]
    ~. \:\:\:\:\:
 \eeqn
We notice that using Eqs.~(\ref{DZS1}) and (\ref{DZgmug5ext}) one has
 \beqn
    \frac{g_0^2}{16 \pi^2} C_F \left[\D^{ext}_{\gamma_\mu \gamma_5} + 
    \D_{\Sigma_1} \right] = Z_A^{(NS)} - 1 ~,
 \eeqn
where $Z_A^{(NS)}$ is given by Eq.~(\ref{Z_A}). 

\indent For completeness, we report also the matching constants for the operators $\bar{\psi}(x) \gamma_\mu \psi(x)$, $\bar{\psi}(x) \gamma_\mu \gamma_5 \psi(x)$ and $\bar{\psi}(x) \sigma_{\mu \nu} \psi(x)$ \cite{Gabrielli}:
 \beqn
    \D_{\gamma_\mu} & = & - 2 + \ln a^2 \mu^2 - F_{0001} + \gamma_E 
    - 16 \pi^2 \int_{-\pi}^\pi \frac{d^4q}{(2 \pi)^4} \Bigl[ - \frac{\D_3}{16 
    \D_1^3} + \frac{\D_4}{\D_1 \D_2^2}  \nn \\ 
    & + & \frac{r^2}{2 \D_2^2} \left( \D_1 \D_5 + 3 \D_3 \right) + r^4 
    \frac{\D_1^2}{\D_2^2} + \frac{r^2 \csw}{\D_2^2} \left( - \D_3 \D_5 + 
    \D_4 \right) \nn \\
    & + & \frac{r^2 \csw^2 \D_3}{2 \D_1 \D_2^2} \left( \D_3 \D_5 - \D_4 
    \right) \Bigl] \\
    \D_{\gamma_\mu \gamma_5} & = & - 2 + \ln a^2 \mu^2 - F_{0001} + \gamma_E 
    - 16 \pi^2 \int_{-\pi}^\pi \frac{d^4q}{(2 \pi)^4} \Bigl[ - \frac{\D_3}{16 
    \D_1^3} + \frac{\D_4}{\D_1 \D_2^2}  \nn \\ 
    & + & \frac{r^2}{2 \D_2^2} \left( - \D_1 \D_5 + 5 \D_3 \right) + r^4 
    \frac{\D_1^2}{\D_2^2} + \frac{r^2 \csw}{\D_2^2} \left( \D_3 \D_5 - \D_4 
    \right) \nn \\
    & + & \frac{r^2 \csw^2 \D_3}{2 \D_1 \D_2^2} \left( \D_4 - \D_3 \D_5 
    \right) \Bigl] \\
    \D_{\sigma_{\mu \nu}} & = & - 16 \pi^2 \int_{-\pi}^\pi \frac{d^4q}{(2 
    \pi)^4} \Bigl[ \frac{1}{3 \D_1 \D_2^2} \Bigl( \D_1 \D_3 - 4 (\D_3 - 
    \D_4) \Bigl) + 2 r^2 \frac{\D_3}{\D_2^2} + r^4 \frac{\D_1^1}{\D_2^2}
    \nn \\
    & + & \frac{2 r^2 \csw}{3 \D_2^2} \left( \D_4 - \D_3 \D_5 \right) + 
    \frac{r^4 \csw^2 \D_1}{3 \D_2^2} \left( \D_4 - \D_3 \D_5 \right) \Bigl]
 \eeqn

\indent The numerical values of the matching constants $\D_I$, $\D_{\gamma_5}$, $\D_{\gamma_\mu}$, $\D_{\gamma_\mu \gamma_5}$, $\D_{\sigma_{\mu \nu}}$ and $\D^{ext}_{\gamma_\mu \gamma_5}$, obtained for various values of $r$, are collected in Tables \ref{tab:O_csw=0} and \ref{tab:O_csw=1}. Our results for the local operators have been explicitly checked against the corresponding ones of Refs.~\cite{Gabrielli,Zhang1,Capitani}, while those for $\D^{ext}_{\gamma_\mu \gamma_5}$ at $c_{SW} = 0$ coincide with the ones of Ref.~\cite{Zhang2}.

\begin{table}[htb]

\vspace{0.5cm}

\begin{center}

\begin{tabular}{|*{7}{p{1.6cm}}|}
\hline \hline
$r$ & $\D_I$ & $\D_{\gamma_5}$ & $\D_{\gamma_\mu}$ & $\D_{\gamma_\mu \gamma_5}$ & $\D_{\sigma_{\mu \nu}}$ & $\D^{ext}_{\gamma_\mu \gamma_5}$ \\
\hline
$0.0$  &-33.16  &-33.16        &-8.74          &-8.74         &0.74          &6.04                           \\
$0.2$  &-18.63  &-26.72        &-8.97          &-4.92         &-0.37          &3.52                           \\
$0.4$  &-8.34   &-19.75        &-8.76          &-3.05         &-1.86          &3.09                           \\
$0.6$  &-3.85   &-15.10        &-8.36          &-2.73         &-2.90          &3.47                           \\
$0.8$  &-1.55   &-11.98        &-8.02          &-2.80         &-3.63          &3.88                           \\
$1.0$  &-0.10   &-9.74         &-7.76          &-2.94         &-4.17          &4.19                           \\
\hline \hline

\end{tabular}

\end{center}

\caption{Numerical results for various $\D_{\mc{O}}$ for different values of $r$ and with $\csw=0$.}

\label{tab:O_csw=0}

\end{table}

\begin{table}[htb]

\vspace{0.5cm}

\begin{center}

\begin{tabular}{|*{7}{p{1.6cm}}|}
\hline \hline
$r$ & $\D_I$ & $\D_{\gamma_5}$ & $\D_{\gamma_\mu}$ & $\D_{\gamma_\mu \gamma_5}$ & $\D_{\sigma_{\mu \nu}}$ & $\D^{ext}_{\gamma_\mu \gamma_5}$\\
\hline
$0.0$  &-33.16  &-33.16        &-8.74          &-8.74         &0.74          &6.04                          \\
$0.2$  &-21.80  &-27.09        &-8.27          &-5.62         &0.23          &3.85                          \\
$0.4$  &-14.69  &-20.89        &-7.45          &-4.35         &-0.61          &3.27                          \\
$0.6$  &-11.94  &-17.05        &-6.82          &-4.27         &-1.23          &3.29                          \\
$0.8$  &-10.73  &-14.70        &-6.40          &-4.42         &-1.64          &3.38                          \\
$1.0$  &-10.10  &-13.17        &-6.12          &-4.59         &-1.93          &3.46                          \\
\hline \hline

\end{tabular}

\end{center}

\caption{The same as in Table \ref{tab:O_csw=0} but with $\csw=1$.}

\label{tab:O_csw=1}

\vspace{0.5cm}

\end{table}

\indent In order to get a further independent check of the expression (\ref{C}) for the constant $C$ one can consider the relation
 \beqn
    m_R = Z_m \left( m_0 - \frac{g_0^2}{16 \pi^2} C_F \frac{\Sigma_0}{a} 
    \right) = \frac{ m_0 - \bar{M}^{(NS)}}{Z_{\gamma_5}} ~,
    \label{!!!}
 \eeqn
where the first equality follows from the definition (\ref{defDZm}), while the second is a consequence of Eq.~(\ref{check}) and of the definition (\ref{defZmatchO}). Now, using the one-loop expressions
 \beqn 
    Z_m & = & 1-\frac{g_0^2}{16\pi^2}C_F \Bigl(\D_{\Sigma_1} + 
    \D_{\Sigma_2} \Bigl) ~, \nn \\ 
    \bar{M}^{(NS)} & = & \frac{g_0^2}{16 \pi^2} C_F  ( \frac{\Sigma_0}{a} + 
    m_0\,C) ~, \nn \\ 
    Z_{\gamma_5} & = & 1 + \frac{g_0^2}{16\pi^2}C_F \Bigl(\D_{\gamma_5} + 
    \D_{\Sigma_1} \Bigl) ~,
 \eeqn 
one easily finds
 \beqn
    C = - \D \Sigma_2 - \D \gamma_5 ~ ,
 \eeqn
which can be checked analytically using Eqs.~(\ref{C}), (\ref{DZS2}) and (\ref{DZg5}).

\section{Singlet axial $WI$'s} \label{section:SWIs}

\indent Let us now turn to the singlet $WI$ (\ref{SWIa})-(\ref{SWIb}). Because of the anomaly, the operator $X^0 + X_C^0$, representing the chiral variation of the Wilson and Clover terms, can be written as \cite{Bochicchio}
 \beqn 
    X^0(x) + X_C^0(x) & = & \bar{X}^0(x) + \bar{X}_C^0(x) - 2 \bar{M}^{(S)} 
    \bar{\psi}(x) \gamma_5 \psi(x) + \left( 1 - Z_A^{(S)} \right) \de_\mu 
    A_\mu(x) \nn \\
    & + & 2 N_F \frac{g_0^2}{32 \pi^2} Z_{\GGt} \GGt_{sub}(x) ~,
    \label{Xmix}
 \eeqn 
where, again, $\bar{X}^0$ and $\bar{X}_C^0$ indicate dimension-5 operators vanishing in the continuum limit in all matrix elements involving elementary operator insertions, and the symbol $\GGt_{sub}$ indicates a lattice discretization of the corresponding continuum operator $\GGt$. In Eq.~(\ref{Xmix}) $\overline{M}^{(S)}$ is the additive mass-renormalization constant, while $Z_A^{(S)}$ and $Z_{\GGt}$ are multiplicative constants in the singlet channel.

\indent Before addressing the calculation of the renormalization constants, we point out that in Eq.~(\ref{Xmix}) $\GGt_{sub}$ should be defined in such a way to avoid ambiguities with the pseudoscalar density. Indeed, let us consider the commonly used discretization $\GGt_L$ obtained through the use of the symmetric plaquette, viz.~\cite{Mandula}
 \beqn 
    \GGt_L(x) \equiv \varepsilon_{\mu \nu \rho \sigma} Tr(P_{\mu \nu}(x) 
    P_{\rho \sigma}(x)) ~,
    \label{GGtdef}
 \eeqn 
where $P_{\mu \nu}$ is defined in Eq.~(\ref{Pmunu}) and the sum over the greek indices is understood. Expanding the definition (\ref{GGtdef}) in powers of $a$, one gets
 \beqn 
    \GGt_L(x) = \GGt(x) + a^2 O_6(x)
    \label{GGt_L}
 \eeqn 
where $\GGt = 1/2 \varepsilon_{\mu \nu \rho \sigma} G^a_{\mu \nu} G^a_{\rho \sigma}$ and $O_6$ denotes a dimension-6 operator. Since with Wilson fermions chiral symmetry is broken, there is no guarantee that the operator $O_6(x)$ does not mix with the pseudoscalar density. In case of mixing one expects by dimensional reasons a (cubic) divergence, leading to a linear-divergent mixing of $\GGt_L$ with the pseudoscalar density. Such a divergence has to be subtracted and therefore the operator $\GGt_{sub}$, appearing in Eq.~(\ref{Xmix}), should be related to $\GGt_L$ by
 \beqn
      \GGt_{sub}(x) = \GGt_L(x) - \D \overline{M}_L \bar{\psi}(x) \gamma_5 
      \psi(x)
      \label{deltaML}
 \eeqn
where $\D \overline{M}_L$ contains the $1 / a$ divergence and is in general dependent on the particular discretization $\GGt_L$. Up to now, a non-perturbative definition of $\D \overline{M}_L$ is not known, and we anticipate here that at one-loop order we find no mixing, i.e. $\D \overline{M}_L = O(g_0^4)$. However a non-perturbative definition of $\D \overline{M}_L$ is mandatory, particularly outside the chiral limit. To this end we require that $\GGt_{sub}$ should be a chiral singlet.

\indent First of all our requirement on $\GGt_{sub}$ is needed to guarantee the proportionality between $(m_0 - \overline{M}^{(NS)})$ and $(m_0 - \overline{M}^{(S)})$, as it is demonstrated in Ref.~\cite{Testa}. This implies that: ~ i) the chiral point is the same in the singlet and non-singlet channels; and ~ ii) the renormalized mass is the same in the two channels and it can be therefore identified with the continuum renormalized mass. The latter point follows from a suitable choice of the renormalization constants for the singlet and octet pseudoscalar densities, which in turn arises from the transformation properties of both scalar and pseudoscalar densities as members of the same chiral multiplet.

\indent Let us now consider the non-singlet $WI$ and the case $\mc{O} = S^a(y_1, y_2) ~ \GGt_{sub}(z)$, where $S^a(y_1, y_2) = \bar{\psi}(y_1) \lambda^a \psi(y_2) / 2$ with $y_1 \neq y_2$; one has
 \beqn 
    \D_\mu^x \< \hat{A}_\mu^a(x) S^a(y_1, y_2) ~ \GGt_{sub}(z) \>  & = & 
    2 m_L \< P_5^a(x) \, S^a(y_1, y_2) ~ \GGt_{sub}(z) \> \nn \\
    & - & C_F \left[ \delta(x - y_1) + \delta(x - y_2) \right] ~ \< 
    P_5(y_1, y_2) ~ \GGt_{sub}(z) \> \nn \\
    & + & \< \left[ \bar{X}^a(x) + \bar{X}_C^a(x) \right ] S^a(y_1, y_2) ~ 
    \GGt_{sub}(z) \> , \qquad \qquad \qquad
    \label{NSWI+GGt}
 \eeqn 
where $P_5^a(x) = \bar{\psi}(x) \lambda^a \gamma_5 \psi(x) / 2$, $P_5(y_1, y_2) = \bar{\psi}(y_1) \gamma_5 \psi(y_2)$, $\hat{A}_\mu^a(x) = Z_A^{(NS)} A_\mu^a(x)$, $m_L = m_0 - \overline{M}^{(NS)}$ and the sum over the flavor index $a$ ($a = 1, ..., 8$) is understood. In Eq.~(\ref{NSWI+GGt}) we have considered that $\delta \GGt_{sub}(z) / \delta(i \alpha^a(x))$ $= 0$, because $\GGt_{sub}$ is assumed to be a chiral singlet. After integration over the whole space-time in $x$ and $z$\footnote{The integration over $x$ is used to cancel out the l.h.s.~of Eq.~(\ref{NSWI+GGt}), including possible contact terms between $\hat{A}_\mu^a(x)$ and $\GGt_{sub}(z)$. The integration over $z$ allows to get rid of the mixing of $\GGt_{sub}(z)$ with the singlet $\de_\mu A_\mu(x)$ [see below Eq.~(\ref{mixing})]. Note that such a mixing does not change the property of the renormalized $\GGt_R$ to be a chiral singlet.} one gets
 \beqn 
   0 & = & m_L \< \int d^4x ~ d^4z ~ P_5^a(x) \, S^a(y_1, y_2) ~ \GGt_{sub}(z) 
   \> - C_F \< P_5(y_1, y_2) \int d^4z ~ \GGt_{sub}(z) \> \nn \\
   & + & {1 \over 2} \< \int d^4x ~ d^4z ~ \left[ \bar{X}^a(x) + \bar{X}_C^a(x) 
   \right ] S^a(y_1, y_2) ~ \GGt_{sub}(z) \>~ .
   \label{prescription}
 \eeqn
We have now to discuss the possible presence of contact terms which may arise in the first and third terms of the r.h.s.~of Eq.~(\ref{prescription}) when $x \approx z$. Note that the operator $S^a(y_1, y_2)$ is a string of elementary operators taken at different space-time points $y_1 \neq y_2$; therefore it cannot generate contact terms when inserted with composite operators. The general structure of the possible contact terms in Eq.~(\ref{prescription}) can be derived using the results of Ref.~\cite{MMRT}, where the method of functional integral with the generalized mass term, $\sigma^a \bar{\psi} \lambda^a \psi / 2 + i \pi^a \bar{\psi} \lambda^a \gamma_5 \psi / 2$, was developed to derive $WI$'s. The only possible contact terms should be as follows
 \beqn
    & & \int d^4x ~ d^4z ~ P_5^a(x) ~ S^a(y_1, y_2) ~ \mc{O'}(z) ~, \nn \\
    & & \int d^4x ~ d^4z ~ \left[ \bar{X}^a(x) + \bar{X}_C^a(x) \right ] 
    S^a(y_1, y_2) ~ \mc{O'}(z) ~, \nn \\
    & & \int d^4x ~ d^4z ~ S^a(y_1, y_2) ~ {\delta \mc{O'}(z) \over \delta 
    \pi^b(x)} ~ \delta_5^a \pi^b(x) ~ ,
    \label{CTs}
 \eeqn
where $\delta_5^a \pi^b(x)$ is the chiral variation of the mass field $\pi^b(x)$ and $\mc{O'}(z)$ is an operator with the same transformation properties and dimension of $\GGt_{sub}(z)$, i.e. a pseudo-scalar, chiral-singlet dimension-4 operator, build up with the fields $\bar{\psi}(z)$, $\psi(z)$, $\sigma^a(z)$ and $\pi^a(z)$ \cite{MMRT}. Since $\mc{O'}(z)$ should be a chiral singlet, the third form of the contact terms (\ref{CTs}) is vanishing. Moreover, the only dimension-4, chiral singlet operator allowed is just the generalized mass term, i.e.: $\mc{O'} = \sigma^a \bar{\psi} \lambda^a \psi / 2 + i \pi^a \bar{\psi} \lambda^a \gamma_5 \psi / 2$, which however is not pseudoscalar. Thus, we conclude that no contact terms are present in Eq.~(\ref{prescription}). Finally, according to Refs.~\cite{Bochicchio,MMRT}  in the continuum limit the third term in the l.h.s. of Eq.~(\ref{prescription}) vanishes and therefore it will be disregarded.

\indent Substituting $\GGt_{sub}(z)$, defined in Eq.~(\ref{deltaML}), into Eq.~(\ref{prescription}) one gets
 \beqn
    \D \overline{M}_L = {- C_F \< \int d^4z ~ P_5(y_1, y_2) ~ \GGt_L(z) \>
     + m_L \< \int d^4x ~ d^4z ~ P_5^a(x) ~ S^a(y_1, y_2) ~ \GGt_L(z) \> \over 
     - C_F \< \int d^4z ~ P_5(y_1, y_2) ~ P_5(z) \> + m_L \< \int d^4x ~ d^4z ~ 
     P_5^a(x) ~ S^a(y_1, y_2) ~ P_5(z) \> } ~.
     \label{solution}
 \eeqn
Note that our non-perturbative definition of $\D \overline{M}_L$ is gauge-invariant because the loss of gauge invariance is given by a factor which is the same in the numerator and in the denominator of the r.h.s. of Eq.~(\ref{solution}).

\indent Let us now address the calculation of the renormalization constants $\overline{M}^{(S)}$, $Z_A^{(S)}$ and $Z_{\GGt}$ relevant to the singlet $WI$. Substituting Eq.~(\ref{Xmix}) into the $WI$ (\ref{SWIa})-(\ref{SWIb}), one gets
 \beqn 
    \D_\mu^x \< Z_A^{(S)} A_\mu(x) \, \mc{O} \> & = & 2 (m_0 - \bar{M}^{(S)}) 
    \< \bar{\psi}(x) \gamma_5 \psi(x) \, \mc{O} \> \nn \\
    & + & 2 N_F \frac{g_0^2}{32 \pi^2} \< Z_{\GGt} \GGt_{sub}(x) \, \mc{O} \> - 
    \< {\delta \mathcal{O} \over \delta(i \alpha^a(x))} \> ~,
 \eeqn 
where we have dropped the term involving $\bar{X}^0 + \bar{X}_C^0$. Because of its anomalous non-conservation the singlet current $A_\mu$ suffers a logarithmically divergent renormalization \cite{Bardeen}. A mixing between $Z_A^{(S)} A_\mu$ and $Z_{\GGt} ~ \GGt_{sub}$ is needed to obtain finite operators \cite{Testa}
 \beqn 
    A_\mu^R(x) & = & Z_A^{(S)} A_\mu(x) - (g_0^2 Z_C) Z_A^{(S)} A_\mu(x) ~, 
    \nn \\
    2 N_F \frac{1}{32 \pi^2} \GGt_R(x) & = & 2 N_F \frac{1}{32 \pi^2} 
    Z_{\GGt} \GGt_{sub}(x) - Z_C Z_A^{(S)} \de_\mu A_\mu(x) ~,
 \label{mixing}
 \eeqn 
where we stress that $\GGt_{sub}(x) = \GGt_L(x) - \D \overline{M}_L \bar{\psi}(x) \gamma_5 \psi(x)$. The above redefinition can be performed into Eq.~(\ref{Xmix}) by adding and subtracting the counterterm $\left[ Z_C Z_A^{(S)} \cdot \de_\mu A_\mu(x) \right]$ \cite{EDM}. 

\indent We have evaluated at one loop level all the constants appearing in Eq.~(\ref{mixing}). The constant $Z_A^{(S)}$ can be obtained by evaluating $\< X^0(x) + X_C^0(x) \, \psi(y) \bar{\psi}(z)\>$: to $O(g_0^2)$ its value is identical to that of $Z_A^{(NS)}$, Eq.~(\ref{Z_A}), while beyond the one-loop order the two constants will differ because of the contributions (in the singlet channel) coming from fermion loops. The same is true for $\bar{M}^{(S)}$, which therefore at one-loop is given by Eq.~(\ref{Mbar}), as well as for the renormalization constants of the pseudoscalar density operators in both the singlet and non-singlet channels. This implies that at one-loop the renormalized masses $(m_0 - \overline{M}^{(S)}) / Z_{\gamma_5}^{(S)}$ and $(m_0 - \overline{M}^{(NS)}) / Z_{\gamma_5}^{(NS)}$ coincide and match the corresponding renormalized mass in the continuum. We stress that the equality between the renormalized masses in the singlet and non-singlet channels is a more general result, valid at any order of the perturbation theory and resulting from the property of $\GGt_{sub}$ to be a chiral singlet, as illustrated in Ref.~\cite{Testa}.

\indent The mixing between $X^0(x) + X_C^0(x)$ and $\GGt_{sub}(x)$ can be computed at one loop by evaluating the correlator $\< \left[ X^0(x) + X_C^0(x) \right] \, G^b_{\beta}(y) G^c_{\gamma}(z) \>$. This was carried out in Ref.~\cite{KarstenSmit}, where it was shown that the apparent dependence of the correlator on $r$ actually disappears as far as $r \neq 0$, and that the tree level expression $2 N_F g_0^2 /32 \pi^2 \, \GGt$ is reproduced exactly. Thus, from Eq.~(\ref{Xmix}) one has $Z_{\GGt} = 1 + O(g_0^2)$, and, since the logarithmic divergence in the singlet axial current manifests itself at the two-loop level, it follows that $g_0^2 Z_C = O(g_0^4)$ and consequently $Z_C = O(g_0^2)$. 

\indent The same considerations hold as well for the matching constants between the continuum and the lattice. For $Z_{\GGt}$ and $Z_C$ appearing in Eq.~(\ref{mixing}), we can write
 \beqn 
    Z_{\GGt} & = & 1 + g_0^2 z^{(2)}_{gg} + O(g_0^4) ~, \nn \\
    Z_C & = & \frac{g_0^2}{16\pi^2}C_F \left( z^{(2)}_{g \psi} \ln (a^2 
    \mu^2) + \tilde{z}^{(2)}_{g \psi} \right) + O(g_0^4) ~.
    \label{Zetas}
 \eeqn 
The coefficient $z^{(2)}_{gg}$ can be computed at one-loop in a pure gauge theory (see Ref.~\cite{Vicari} and references therein quoted), and at this order it is unaffected by the presence of the Clover term in the action\footnote{Beyond $O(g_0^2)$ the presence of the Clover term in the action will, in general, affect also $Z_{\GGt}$.}. One obtains \cite{Vicari}
 \beqn 
    z^{(2)}_{gg} = N \left( -\frac{1}{4 N^2} + Z_{0000} + \frac{1}{8} + 
    \frac{1}{2 \pi^2} \right)
 \eeqn 
where $N$ is the dimension of the gauge group and $Z_{0000} = 0.15493$. The constants $z^{(2)}_{g \psi}$ and $\tilde{z}^{(2)}_{g \psi}$ can be computed by evaluating, in the continuum and on the lattice, the correlator $\< \GGt(x) \psi(y) \bar{\psi}(z)\>$, with amputated external quark propagators. On the lattice $\GGt$ should be $\GGt_{sub}$ given by Eq. (\ref{deltaML}). However, we have explicitly checked that the operator $\GGt_L$ does not mix at one-loop with the pseudoscalar density and, therefore, in our leading-order approximation there is no difference between $\GGt_{sub}$ and $\GGt_L$. Thus, on the lattice we have to compute the following diagrams:
 \beqn 
    FT \< \GGt_L(x) \, \psi(y) \bar{\psi}(z)  \>^{(2)}_{amp} &=& \quad \:\: 
    \parbox{5mm}{
    \begin{fmfgraph*}(15,15)
      \fmfleft{i1,i2} \fmfright{o1,o2}
      \fmf{plain,tension=3}{i1,v0} 
      \fmfv{decor.shape=cross,decor.angle=55}{v0}
      \fmf{plain,tension=3}{v0,v1} 
      \fmf{photon}{v1,V1}
      \fmf{phantom}{i2,V1} 
      \fmf{phantom}{V1,o2}
      \fmf{photon}{V1,v2}
      \fmf{plain,tension=3}{v2,v3} 
      \fmfv{decor.shape=cross,decor.angle=-55}{v3}
      \fmf{plain,tension=3}{v3,o1}
      \fmffreeze
      \fmf{quark,right=0.2}{v1,v2}
      \fmfv{decor.shape=square,decor.size=4.thin,decor.filled=full}{V1} 
      \fmflabel{$\GGt_L$}{V1}
    \end{fmfgraph*}} \qquad \qquad \qquad \quad + 
    \parbox{5mm}{
    \begin{fmfgraph*}(15,15)
      \fmfleft{i1,i2,i3} \fmfright{o1,o2,o3}
      \fmf{quark,tension=2}{i1,v0} 
      \fmfv{decor.shape=cross,decor.angle=40}{v0}
      \fmf{plain,tension=2}{v0,V1}
      \fmf{phantom}{i2,V1} 
      \fmf{phantom}{V1,o2}
      \fmf{plain,tension=2}{V1,v2} 
      \fmfv{decor.shape=cross,decor.angle=-40}{v2}
      \fmf{quark,tension=2}{v2,o1} 
      \fmffreeze
      \fmf{phantom}{i3,V2} 
      \fmf{phantom}{V2,o3}
      \fmf{photon,right=1}{V1,V2} 
      \fmf{photon,right=1}{V2,V1} 
      \fmf{phantom,tension=0.5}{V2,V1}
      \fmfv{decor.shape=square,decor.size=4.thin,decor.filled=full}{V2} 
      \fmflabel{$\GGt_L$}{V2}
    \end{fmfgraph*} } \qquad \qquad \qquad
\eeqn 
where the $\psi \psi g$ and $\psi \psi g g$ vertices include the contributions of both the Wilson and Clover action. The second diagram gives a null contribution, since the insertion of $\GGt$ contracted with two gluon lines is antisymmetric in the Lorentz indices of the gluons (via the $\varepsilon$-tensor) and symmetric in their color indices, whereas the $\psi \psi g g$ is symmetric in the Lorentz indices for the Wilson case and antisymmetric in the color matrices for the Clover case (see the Feynman rules in the Appendix). The first diagram displays a log divergence as $a \ra 0$, in analogy with the continuum calculation. The difference between the two schemes gives the following result
 \beqn 
    z^{(2)}_{g \psi} & = & -~6~, \nn \\ [0.5cm]
    \tilde{z}^{(2)}_{g \psi} & = & 22 - 6(\gamma_E - F_{0001}) - 16 \pi^2 
    \int \frac{d^4 q}{(2 \pi)^4} ~ \Bigl[ \frac{3 \D_3}{2 \D_1^2} \left( 
    \frac{1}{4 \D_1} - \frac{1}{\D_2}\right) \nn \\ 
    & + & \frac{1}{4 \D_1^2 \D_2} \left( 10 \D_3 \D_4 - 30 \D_3 - 19 \D_3^2 
    + 64 \D_4 - 46 \D_7 + 12 \D_8 \right) \nn \\
    & + & \frac{1}{4 \D_1 \D_2} \left( 45 \D_3 + 6 \D_3^2 - 40 \D_4 + 12 
    \D_7 \right) + \frac{1}{4 \D_2} \left( 2 \D_1 \D_3 - 17 \D_3 + 6 \D_4 
    \right) \nn \\
    & + & r^2 c_{SW} \Bigl[ \frac{1}{2 \D_1 \D_2} \Bigl(  8 \D_3 \D_4 - 
    3 \D_3 \D_5^2 + 3 \D_3 \D_6 - 4 \D_3^2 \D_5 + 10 \D_4 \D_5 + 2 \D_4 
    \D_5^2 \nn \\ 
    & - & 2 \D_4 \D_6 - 8 \D_5 \D_7 - 14 \D_7 + 12 \D_8 \Bigl) + 
    \frac{1}{\D_2} \left( \D_3 \D_5^2 - \D_3 \D_6 - 2 \D_4 \D_5 +2 \D_7 
    \right) \Bigl] \nn \\
    & + & r^2 c^2_{SW} \Bigl[ \frac{1}{4 \D_1^2 \D_2} \Bigl( - 10 \D_3 \D_4 
    \D_5 - 2 \D_3 \D_4 \D_5^2 + 2 \D_3 \D_4 \D_6 + 8 \D_3 \D_5 \D_7 \nn \\
    & + & 14 \D_3 \D_7 - 12 \D_3 \D_8 - 8 \D_3^2 \D_4 + 3 \D_3^2 \D_5^2 
    - 3 \D_3^2 \D_6 + 4 \D_3^3 \D_5 \Bigl) \nn \\
    & + & \frac{1}{2 \D_1 \D_2} \left(  2 \D_3 \D_4 \D_5 - 2 \D_3 \D_7 - 
    \D_3^2 \D_5^2  + \D_3^2 \D_6 \right) \Bigl] ~.
    \label{zgpsi}
 \eeqn 

\indent The numerical values of the constant $\tilde{z}^{(2)}_{g \psi}$ obtained for various values of $r$ at $\csw = 0$ and $\csw = 1$ are reported in Table \ref{tab:zgpsi}.

\begin{table}[htb]

\begin{center}

\begin{tabular}{|*{3}{p{1.92cm}}|}
\hline \hline
$r$ &  $ \tilde{z}^{(2)}_{g \psi} $   & $ \tilde{z}^{(2)}_{g \psi} $\\
    &  $\scst \csw=0$   & $\scst \csw=1$ \\
\hline
$0.0$  &19.82    &19.82              \\  
$0.2$  &19.72    &19.93              \\
$0.4$  &19.11    &19.98              \\
$0.6$  &18.11    &19.94              \\
$0.8$  &16.95    &19.87              \\
$1.0$  &15.77    &19.81              \\
\hline \hline

\end{tabular}

\end{center}

\caption{Numerical results for $\tilde{z}^{(2)}_{g \psi}$ [see Eq.~(\ref{zgpsi})] for different values of $r$ and $\csw$.}

\label{tab:zgpsi}

\end{table}

\subsection{Composite insertions of the topological charge and the neutron $EDM$ \label{subsection:EDM}}

\indent In this subsection we take the opportunity to briefly address a  separate issue concerning the composite insertion of the topological charge relevant for a lattice evaluation of the neutron $EDM$ induced by the strong $\theta$-term (\ref{eq:theta}). Following Ref.~\cite{EDM} the standard definition of the neutron $EDM$, $\vec{d}_N$, involves the insertion of the topological charge $(g^2 / 32 \pi^2) \int d^4x ~ \GGt(x)$ in the presence of the charge density operator $J_0$. Treating the $\theta$-term as a perturbation at first order, one has
 \beqn
    \vec{d}_N \equiv -i \theta {g^2 \over 32 \pi^2} \int d^3y ~ \vec{y} ~
    _0\langle N | J_0(y) \left[ \int d^4x ~ \GGt(x) \right] | N \rangle_0 ~, 
    \label{eq:EDM}
 \eeqn
where $| N \rangle_0$ is a shorthand for $| N \rangle_{\theta = 0}$. In case of three flavors with non-degenerate masses a complete diagrammatic analysis was performed in Ref.~\cite{EDM} showing how the axial anomaly governs the replacement of the topological charge operator with well-defined insertions of the flavor-singlet pseudoscalar density\footnote{We point out that the final result of Ref.~\cite{EDM} crucially depends on the equality between the renormalized masses in the singlet and non-singlet $WI$'s. We stress once more that such an equality follows from the property of $\GGt_{sub}$ [see Eq.~(\ref{deltaML})] to be a chiral singlet (cf.~Ref.~\cite{Testa}).}.

\indent Thus the question is the possible presence of contact terms in the composite insertion of the topological charge operator with the electromagnetic ($e.m.$) current operator, which would lead to ambiguities in the numerical evaluation of Eq.~(\ref{eq:EDM}). The contact terms in Eq.~(\ref{eq:EDM}) should be of the form $O_{\mu =0}^P(x) \cdot \delta^{(4)}(x - y)$, where $O_\mu^P$ is a local operator of dimension-3, which transforms as a non-singlet pseudo-vector and is conserved. The last property derives from the fact that the $e.m.$ current is conserved both with and without the $\theta$-term in the $QCD$ action at any value of the parameter $\theta$. The non-singlet nature of $O_\mu^P$ is related to the non-singlet nature of the $e.m.$ current operator. The only candidate for $O_\mu^P$ is the non-singlet axial current which however is not conserved (outside the chiral limit).

\indent The absence of contact terms in Eq.~(\ref{eq:EDM}) can be derived also using the generalized mass insertion method of Ref.~\cite{MMRT} applied to a vector $WI$ with the operator $\mc{O}$ given by $\GGt$.

\section{Conclusions \label{section:conclusions}}

\indent A complete one-loop calculation of the renormalization constants appearing in both singlet and non-singlet axial Ward identities using Wilson fermions with the Clover $O(a)$-improvement of the action has been performed. Our calculations include: ~ 1) the contributions arising from the Clover term of the action; ~ 2) the complete mixing of the gluon operator $\GGt$ with the divergence of the singlet axial current; ~ 3) the use of both local and extended definitions of the fermionic bilinear operators.

\indent In the singlet channel a definition of the gluon operator $\GGt$ on the lattice outside the chiral limit has been proposed. Our definition takes into account the possible power-divergent mixing with the pseudoscalar density, generated by the breaking of chiral symmetry. No mixing has been found at one-loop order and a non-perturbative definition of the mixing constant has been developed.

\indent Finally, the renormalization properties of the composite insertion of the topological charge operator $\int d^4x ~ \GGt(x)$ relevant for the lattice calculation of the neutron electric dipole moment have been discussed, showing that no contact terms arise when the topological charge is inserted with a conserved current.

\section*{Acknowledgments}

\indent The authors gratefully acknowledge M.~Testa for many useful discussions, which have been essential for clarifying the non-perturbative definition of the mixing between the gluon operator $\GGt$ on the lattice and the singlet pseudoscalar density. We warmly thank V.~Lubicz and G.~Martinelli for many useful comments and a critical reading of the manuscript.

\section*{Appendix: Feynman rules}

\appendix

\label{feyrules}

\indent The Wilson action is given by
 \beqn
   S_F & = & a^4 \sum_{x} \Bigl\{ -\frac{1}{2a} \sum_{\mu}[\bar{\psi}(x)(r - 
   \gamma_\mu) U_\mu(x) \psi(x + \mu) \nn \\ 
   & + & \bar{\psi}(x + \mu)(r + \gamma_\mu) U_\mu^\dagger(x)\psi(x)] + 
   \bar{\psi}(x) \Bigl( m_0 + \frac{4r}{a} \Bigl) \psi(x) \Bigl\} ~,
 \eeqn
with the following notation $\{\gamma_\mu ,\gamma_\nu \} = 2 \delta_{\mu \nu}$, $\sigma_{\mu \nu} = [\gamma_\mu , \gamma_\nu] / 2$, $U_\mu(x) = \exp [i g_0 a G_\mu(x)]$, $G_\mu = G_\mu^a t^a$ and $Tr (t^a t^b) = \delta_{ab} / 2$.

\indent The action is improved to $O(a)$ via the Clover term \cite{Heatlie}
 \beqn 
    S_C = - a^4 \sum_{x} \sum_{\mu, \nu} \: \csw \: \frac{i g_0 a r}{4} 
    \bar{\psi}(x)\sigma_{\mu \nu} P_{\mu \nu}(x) \psi(x) ~,
 \eeqn 
with $P_{\mu \nu}$ being the usual lattice definition of the field-strength tensor $G_{\mu \nu}$ \cite{Gabrielli}
 \beqn 
    P_{\mu \nu}(x) = \frac{1}{4 a^2} \sum_{i=1}^4 \frac{1}{2 i g_0} (U_i - 
    U_i^\dagger)
 \eeqn 
where the sum is over the four plaquettes in the $\mu$-$\nu$ plane stemming from $x$ and taken in the counterclockwise sense, i.e.
 \beqn 
    \label{Ui}
    U_1 & = & U_\mu(x) ~ U_\nu(x+\mu) ~ U_\mu^\dagger(x+\nu) ~ 
    U_\nu^\dagger(x) ~, 
    \nn \\
    U_2 & = & U_\nu(x) ~ U_\mu^\dagger(x-\mu+\nu) ~ U_\nu^\dagger(x-\mu) ~ 
    U_\mu(x-\mu) ~, \nn \\
    U_3 & = & U_\mu^\dagger(x-\mu) ~ U_\nu^\dagger(x-\mu-\nu) ~
    U_\mu(x-\mu-\nu) ~ U_\nu(x-\nu) ~, \nn \\
    U_4 & = & U_\nu^\dagger(x-\nu) ~ U_\mu(x-\nu) ~ U_\nu(x+\mu-\nu) ~ 
    U_\mu^\dagger(x) ~.
 \eeqn

\section*{Propagators}

 \beqn
    \parbox{15mm}{
    \begin{fmfgraph*}(15,10)
      \fmfleft{i1} \fmflabel{$\mu$}{i1} \fmfdot{i1}
      \fmfright{o1} \fmflabel{$\nu$}{o1} \fmfdot{o1}
      \fmf{photon,label=$k$,label.side=right}{i1,o1}
    \end{fmfgraph*} } \qquad \qquad \qquad \equiv \mc{G}_{\mu \nu}(k) = 
    \frac{\delta_{\mu \nu} - (1-\alpha) \frac{\hat{k}_\mu 
    \hat{k}_\nu}{\hat{k}^2}}{\hat{k}^2}
 \eeqn
with
 \beqn
    \hat{p}_\mu = \frac{2}{a} \sin(\frac{p_\mu a}{2}), \qquad \hat{p}^2 = 
    \sum_{\mu} \hat{p}_\mu^2 ~. \nn
 \eeqn

 \beqn
    \parbox{15mm}{
    \begin{fmfgraph*}(15,10)
      \fmfleft{i1} 
      \fmfdot{i1}
      \fmfright{o1} 
      \fmfdot{o1}
      \fmf{quark,label=$p$,label.side=right}{i1,o1}
    \end{fmfgraph*} } \qquad \qquad \qquad \equiv S_0(p) = \nn
 \eeqn
\vspace{-4mm}
 \beqn
    = \Bigl[\frac{1}{a} \sum_{\lambda} i \gamma_\lambda \sin (p_\lambda a) + 
    \Bigl(m_0 + \frac{2 r}{a} \sum_{\lambda} \sin^2 \frac{p_\lambda a}{2} 
    \Bigl) \Bigl]^{-1}
 \eeqn

\section*{QCD vertices}

\indent The indices $W$ and $C$ refer to the Wilson and Clover action respectively \cite{Gabrielli}:
 \beqn
    \parbox{20mm}{
    \begin{fmfgraph*}(20,15)
      \fmfleft{i1,i2} 
      \fmflabel{$q,j$}{i1}
      \fmfright{o1,o2} 
      \fmflabel{$q',i$}{o1}
      \fmf{phantom}{i2,V2} 
      \fmf{phantom}{V2,o2}
      \fmf{quark}{i1,V1} 
      \fmf{quark}{V1,o1}
      \fmf{photon}{V1,V2} 
      \fmflabel{$\mu,a$}{V2}
      \fmfv{decor.shape=circle,decor.size=8.thin,decor.filled=empty}{V1}
    \end{fmfgraph*}} \qquad \qquad \qquad \equiv V_{\mu a}^W(q,q') = \nn
 \eeqn
\smallskip
 \beqn
    = - g_0 (t^a)_{ij} \Bigl\{ i \gamma_\mu \cos \Bigl((q + q')_\mu 
    \frac{a}{2}\Bigl) + r \sin\Bigl((q+q')_\mu \frac{a}{2}\Bigl) \Bigl\}
 \eeqn
\bigskip
 \beqn
    \parbox{20mm}{
    \begin{fmfgraph*}(20,15)
      \fmfleft{i1,i2} 
      \fmflabel{$q,j$}{i1}
      \fmfright{o1,o2} 
      \fmflabel{$q',i$}{o1}
      \fmf{phantom}{i2,V2} 
      \fmf{phantom}{V2,o2}
      \fmf{quark}{i1,V1} 
      \fmf{quark}{V1,o1}
      \fmf{photon}{V1,V2} 
      \fmflabel{$\mu,a$}{V2}
      \fmfv{decor.shape=circle,decor.size=8.thin,decor.filled=full}{V1}
    \end{fmfgraph*}} \qquad \qquad \qquad \equiv V_{\mu a}^C(q, q') = \nn
 \eeqn
\smallskip
 \beqn
    = - g_0 (t^a)_{ij} \Bigl\{ \frac{r}{2} \sum_{\nu} \sigma_{\mu \nu} 
    \sin \Bigl((q-q')_\nu a\Bigl) \cos \Bigl((q-q')_\mu \frac{a}{2}\Bigl) 
    \Bigl\}
 \eeqn
\bigskip
 \beqn
    \parbox{5mm}{
    \begin{fmfgraph*}(20,15)
      \fmfleft{i1,i2} \fmflabel{$p,j$}{i1} \fmflabel{$\rho,a$}{i2}
      \fmfright{o1,o2} \fmflabel{$p',i$}{o1} \fmflabel{$\sigma,b$}{o2}
      \fmf{photon}{i2,V1} \fmf{photon}{V1,o2}
      \fmf{quark}{i1,V1} \fmf{quark}{V1,o1}
      \fmfv{decor.shape=circle,decor.size=8.thin,decor.filled=empty}{V1}
    \end{fmfgraph*}} \qquad \qquad \qquad \qquad \qquad  =\: V_{\rho a , 
    \sigma b }^W(p,p') \:=\: \nn
 \eeqn
\medskip
 \beqn
    = \frac{a g_0^2}{2} \delta_{\rho \sigma} \{t^a,t^b\}_{ij} \Bigl[ i 
    \gamma_\rho \sin \Bigl( (p+p')_\rho \frac{a}{2} \Bigl) - r \cos \Bigl( 
    (p + p')_\rho \frac{a}{2} \Bigl) \Bigl]
\eeqn

\section*{Extended operators}

\indent For simplicity we report only the Feynman rules for flavor-singlet operators.

\subsection*{Operator $\D_\mu A_\mu(x)$:}

 \beqn
    a \Delta_\mu A_\mu(x) & = & A_\mu(x) - A_\mu(x-\mu) ~, \nn \\
    A_\mu(x) & = & \frac{1}{2}\Bigl[\bar{\psi}(x+\mu) \gamma_\mu \gamma_5 
    U_\mu^\dagger(x) \psi(x) + \bar{\psi}(x) \gamma_\mu \gamma_5 U_\mu(x) 
   \psi(x+\mu) \Bigl]~.
 \eeqn

 \beqn
    \parbox{15mm}{
    \begin{fmfgraph*}(15,10)
      \fmfleft{i1} \fmflabel{$p$}{i1}
      \fmfright{o1} \fmflabel{$p'$}{o1}
      \fmf{quark}{i1,V} \fmf{quark}{V,o1}
      \fmfv{decor.shape=square,decor.size=4.thin,decor.filled=full,label=
      $\Delta_\mu A_\mu$,label.angle=90}{V}
    \end{fmfgraph*} } \qquad \qquad \qquad =\: -\frac{i}{a} \sum_{\mu} 
    \gamma_\mu \gamma_5 \Bigl( \sin(p_\mu ' a) - \sin(p_\mu a) \Bigl) ~,
 \eeqn

 \beqn
    \parbox{20mm}{
    \begin{fmfgraph*}(20,15)
      \fmfleft{i1,i2} \fmflabel{$p,j$}{i1}
      \fmfright{o1,o2} \fmflabel{$p',i$}{o1}
      \fmf{phantom}{i2,V2} \fmf{phantom}{V2,o2}
      \fmf{quark}{i1,V1} \fmf{quark}{V1,o1}
      \fmf{photon}{V1,V3} \fmf{photon}{V3,V2} \fmflabel{$q,\rho,a$}{V2}
      \fmfv{decor.shape=triacross,decor.size=8.thin,decor.angle=110}{V3}
      \fmfv{decor.shape=square,decor.size=4.thin,decor.filled=full,label=
      $\Delta_\mu A_\mu$}{V1}
   \end{fmfgraph*}} \qquad \qquad \qquad = \: \nn 
 \eeqn
\smallskip
 \beqn
    = \frac{g_0}{2} \gamma_\rho \gamma_5 (t^a)_{ij} \sum_{\mu} \Bigl[ e^{i 
    q_\rho \frac{a}{2}} \Bigl( e^{-i p_\mu ' a} - e^{i p_\mu a} \Bigl) + 
    e^{-i q_\rho \frac{a}{2}} \Bigl( e^{i p_\mu ' a} - e^{-i p_\mu a} 
    \Bigl) \Bigl] ~,
 \eeqn
\bigskip
 \beqn
    \parbox{5mm}{
    \begin{fmfgraph*}(20,15)
      \fmfleft{i1,i2} \fmflabel{$p,j$}{i1} \fmflabel{$q,\rho,a$}{i2}
      \fmfright{o1,o2} \fmflabel{$p',i$}{o1} \fmflabel{$q',\sigma,b$}{o2}
      \fmf{photon,tension=2}{i2,Vi}  \fmf{photon,tension=2}{Vi,V1}
      \fmfv{decor.shape=triacross,decor.size=8.thin,decor.angle=38}{Vi}
      \fmf{photon,tension=2}{V1,Vo} \fmf{photon,tension=2}{Vo,o2}
      \fmfv{decor.shape=triacross,decor.size=8.thin,decor.angle=14}{Vo}
      \fmf{quark}{i1,V1} \fmf{quark}{V1,o1}
      \fmfv{decor.shape=square,decor.size=4.thin,decor.filled=full,label=
      $\Delta_\mu A_\mu$,label.angle=90,label.dist=120}{V1}
    \end{fmfgraph*}} \qquad \qquad \qquad \qquad \qquad = \: 
 \eeqn
\smallskip
 \beqn
    & = & - \frac{a g_0^2}{4} \gamma_\rho \gamma_5 \delta_{\rho \sigma} 
    \{ t^a,t^b\}_{ij} \sum_{\mu} \Bigl[ e^{i q_\rho \frac{a}{2}} e^{-i 
    q_\rho' \frac{a}{2}} \Bigl( e^{-i p_\mu' a} + e^{i p _\mu a} \Bigl) 
    \nn \\ 
    & - & e^{-i q_\rho \frac{a}{2}} e^{i q_\rho' \frac{a}{2}} \Bigl( e^{i 
    p_\mu' a} + e^{-i p _\mu a} \Bigl)\Bigl] ~.
\eeqn

\subsection*{Operator $X^0$:}

 \beqn
    X^0(x) & = & - \frac{r}{2a} \sum_{\mu} \Bigl[ \bar{\psi}(x) \gamma_5 
    U_\mu(x) \psi(x+\mu) + \bar{\psi}(x+\mu) \gamma_5 U_\mu^\dagger(x) 
    \psi(x) \nn \\ 
    & + & (x \rightarrow x-\mu) - 4 \bar{\psi}(x) \gamma_5 \psi(x) \Bigl] 
    ~. \: \: \: 
 \eeqn
 \beqn
    \parbox{15mm}{
    \begin{fmfgraph*}(15,10)
      \fmfleft{i1}  \fmflabel{$p$}{i1}
      \fmfright{o1}  \fmflabel{$p'$}{o1}
      \fmf{quark}{i1,V} \fmf{quark}{V,o1}
      \fmfv{decor.shape=square,decor.size=4.thin,decor.filled=full,label= 
      $X^0$,label.angle=90}{V}
    \end{fmfgraph*}} \qquad \qquad \qquad = \: \frac{r}{a} \gamma_5 
    \sum_{\mu} \Bigl( 2 - \cos(p_\mu a) - \cos(p_\mu ' a)\Bigl) ~,
 \eeqn
\vspace{-1.1cm}
 \beqn
    \parbox{20mm}{
    \begin{fmfgraph*}(20,15)
      \fmfleft{i1,i2} \fmflabel{$p,j$}{i1}
      \fmfright{o1,o2} \fmflabel{$p',i$}{o1}
      \fmf{phantom}{i2,V2} \fmf{phantom}{V2,o2}
      \fmf{quark}{i1,V1} \fmf{quark}{V1,o1}
      \fmf{photon}{V1,V3} \fmf{photon}{V3,V2} \fmflabel{$q,\rho,a$}{V2}
      \fmfv{decor.shape=triacross,decor.size=8.thin,decor.angle=110}{V3}
      \fmfv{decor.shape=square,decor.size=4.thin,decor.filled=full,label=
      $X^0$}{V1}
    \end{fmfgraph*}} \qquad \qquad \qquad = \: \nn 
 \eeqn
\smallskip
 \beqn
     = - \frac{i g_0 r}{2} \gamma_5 (t^a)_{ij} \Bigl[ e^{i q_\rho 
     \frac{a}{2}} \Bigl( e^{i p_\rho  a} - e^{-i p_\rho ' a} \Bigl) + e^{-i 
     q_\rho \frac{a}{2}} \Bigl( e^{i p_\rho ' a} - e^{-i p_\rho a} \Bigl) 
     \Bigl] ~, \nn \\
 \eeqn
\bigskip
 \beqn
    \parbox{5mm}{
    \begin{fmfgraph*}(20,15)
      \fmfleft{i1,i2} \fmflabel{$p,j$}{i1} \fmflabel{$q,\rho,a$}{i2}
      \fmfright{o1,o2} \fmflabel{$p',i$}{o1} \fmflabel{$q',\sigma,b$}{o2}
      \fmf{photon,tension=2}{i2,Vi}  \fmf{photon,tension=2}{Vi,V1}
      \fmfv{decor.shape=triacross,decor.size=8.thin,decor.angle=38}{Vi}
      \fmf{photon,tension=2}{V1,Vo} \fmf{photon,tension=2}{Vo,o2}
      \fmfv{decor.shape=triacross,decor.size=8.thin,decor.angle=14}{Vo}
      \fmf{quark}{i1,V1} \fmf{quark}{V1,o1}
      \fmfv{decor.shape=square,decor.size=4.thin,decor.filled=full,label=
      $X^0$,label.angle=90,label.dist=100}{V1}
    \end{fmfgraph*}} \qquad \qquad \qquad \qquad \qquad = \: \nn
 \eeqn
\smallskip
 \beqn
    & = & \frac{g_0^2 a r}{4} \gamma_5 \delta_{\rho \sigma} \{ t^a,t^b\}_{ij} 
    \sum_{\mu} \Bigl[ e^{i q_\rho \frac{a}{2}} e^{-i q_\rho ' \frac{a}{2}} 
    \Bigl( e^{i p_\mu a} + e^{-i p _\mu ' a} \Bigl) \nn \\ & + & e^{-i q_\rho 
    \frac{a}{2}} e^{i q_\rho ' \frac{a}{2}} \Bigl( e^{i p_\mu ' a} + e^{-i 
    p_\mu a} \Bigl)\Bigl] ~.
 \eeqn

\bigskip

\subsection*{Operator $X_C^0$:} 

 \beqn
    X_C^0(x) = - \frac{i g_0 a r}{2} \sum_{\mu \nu} \bar{\psi}(x) \gamma_5 
    \sigma_{\mu \nu} P_{\mu \nu} \psi(x) ~.
 \eeqn
 \beqn
    \parbox{15mm}{
    \begin{fmfgraph*}(15,10)
      \fmfleft{i1}  \fmflabel{$p$}{i1}
      \fmfright{o1}  \fmflabel{$p'$}{o1}
      \fmf{quark}{i1,V} \fmf{quark}{V,o1}
      \fmfv{decor.shape=square,decor.size=4.thin,decor.filled=full,label=
      $X_C^0$,label.angle=90}{V}
    \end{fmfgraph*}} \qquad \qquad \qquad = \emptyset ~,
 \eeqn
 \beqn
    \parbox{20mm}{
    \begin{fmfgraph*}(20,15)
      \fmfleft{i1,i2} \fmflabel{$p,j$}{i1}
      \fmfright{o1,o2} \fmflabel{$p',i$}{o1}
      \fmf{phantom}{i2,V2} \fmf{phantom}{V2,o2}
      \fmf{quark}{i1,V1} \fmf{quark}{V1,o1}
      \fmf{photon}{V1,V3} \fmf{photon}{V3,V2} \fmflabel{$q,\rho,a$}{V2}
      \fmfv{decor.shape=triacross,decor.size=8.thin,decor.angle=110}{V3}
      \fmfv{decor.shape=square,decor.size=4.thin,decor.filled=full,label=
      $X_C^0$}{V1}
    \end{fmfgraph*}} \qquad \qquad \qquad = \: \nn
 \eeqn
\smallskip
 \beqn
    = - g_0 r \gamma_5 (t^a)_{ij} \sum_{\nu} \Bigl[\sigma_{\rho \nu} 
    \sin(q_\nu a) \cos\Bigl(q_\rho \frac{a}{2}\Bigl)\Bigl] ~,
 \eeqn
\bigskip
 \beqn
    \parbox{5mm}
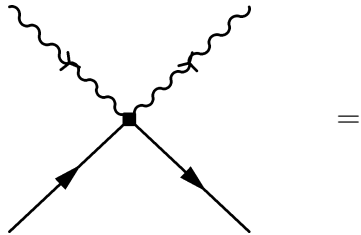
{
    \begin{fmfgraph*}(20,15)
      \fmfleft{i1,i2} \fmflabel{$p,j$}{i1} \fmflabel{$q,\rho,a$}{i2}
      \fmfright{o1,o2} \fmflabel{$p',i$}{o1} \fmflabel{$q',\sigma,b$}{o2}
      \fmf{photon,tension=2}{i2,Vi}  \fmf{photon,tension=2}{Vi,V1}
      \fmfv{decor.shape=triacross,decor.size=8.thin,decor.angle=38}{Vi}
      \fmf{photon,tension=2}{V1,Vo} \fmf{photon,tension=2}{Vo,o2}
      \fmfv{decor.shape=triacross,decor.size=8.thin,decor.angle=14}{Vo}
      \fmf{quark}{i1,V1} \fmf{quark}{V1,o1}
      \fmfv{decor.shape=square,decor.size=4.thin,decor.filled=full,label=
      $X_C^0$,label.angle=90,label.dist=100}{V1}
    \end{fmfgraph*}} \qquad \qquad \qquad \qquad \qquad = \: \nn
 \eeqn
\smallskip
 \beqn
    & = & \frac{g_0^2 a r}{2} \gamma_5 [t^a,t^b]_{ij} \sum_{\nu} \Big\{ 
    \Bigl[ \delta_{\rho \sigma} \sigma_{\rho \nu} \sin \Bigl((q - q')_\rho 
    \frac{a}{2}\Bigl) \Bigl(\sin(q_\nu a) + \sin(q_\nu ' a) \Bigl) \Bigl] 
    \nn \\
    & + & \frac{1}{2} \sigma_{\rho \sigma} \Bigl[ \cos \Bigl((q_\rho - 
    q'_\sigma) \frac{a}{2} \Bigl) - \cos \Bigl((q_\rho-q'_\sigma) 
    \frac{a}{2} - q'_\rho a \Bigl) \nn \\
    & - & \cos \Bigl((q_\rho - q'_\sigma) \frac{a}{2} + q_\sigma a \Bigl) - 
    \cos \Bigl((q_\rho - q'_\sigma) \frac{a}{2} + (q_\sigma - q'_\rho) a 
    \Bigl) \nn \\ 
    & + & \cos \Bigl((q_\rho + q'_\sigma) \frac{a}{2} \Bigl) - \cos 
    \Bigl((q_\rho+q'_\sigma) \frac{a}{2} - q'_\rho a \Bigl) \nn \\ 
    & - & \cos \Bigl((q_\rho+q'_\sigma) \frac{a}{2} - q_\sigma a  \Bigl) - 
    \cos \Bigl((q_\rho + q'_\sigma) \frac{a}{2} - (q_\sigma + q'_\rho) a 
    \Bigl) \Bigl] \Big\} ~.
 \eeqn

\subsection*{Operator $\GGt_L$:} 

 \beqn
    \GGt_L = \varepsilon_{\mu \nu \rho \sigma} Tr( P_{\mu \nu} P_{\rho 
    \sigma}) ~.
\eeqn
\beqn
   \parbox{15mm}{
   \begin{fmfgraph*}(15,10)
      \fmfleft{i1} \fmflabel{$p,\nu,a$}{i1}
      \fmfright{o1} \fmflabel{$r,\mu,b$}{o1}
      \fmf{photon}{i1,Vi} \fmf{photon}{Vi,V} 
      \fmfv{decor.shape=triacross,decor.size=8.thin,decor.angle=-44}{Vi}
      \fmf{photon}{V,Vo} \fmf{photon}{Vo,o1} 
      \fmfv{decor.shape=triacross,decor.size=8.thin,decor.angle=-44}{Vo}
      \fmfv{decor.shape=square,decor.size=4.thin,decor.filled=full,label=
      $\GGt_L$,label.angle=90}{V}
   \end{fmfgraph*} } \qquad \qquad \qquad \qquad \qquad = \nn \\ - 
   \frac{4}{a^2} \delta_{ab} \sum_{\alpha \beta} \varepsilon_{\nu \alpha 
   \mu \beta} \cos(p_\nu \frac{a}{2}) \cos(r_\mu \frac{a}{2}) \sin(p_\alpha 
   a) \sin(r_\beta a)~.
\eeqn

\end{fmffile}

\end{document}